\def\gsim{\;\lower4pt\hbox{${\buildrel\displaystyle >\over\sim}$}\,}
\def\lsim{\;\lower4pt\hbox{${\buildrel\displaystyle <\over\sim}$}\,}

\documentclass[usenatbib]{mn2e}
\usepackage{graphicx}

\newcommand\E[1]{\times10^{#1}}
\newcommand\un[1]{{\,\rm #1}}
\newcommand\rs[1]{_\mathrm{#1}}
\newcommand\g{$\gamma$}
\newcommand\ApJ{ApJ}

\title[Tools for non-thermal images of SNRs]{Radio, X-ray and gamma-ray surface brightness profiles as powerful diagnostic tools for non-thermal SNR shells}
\author[Petruk O. et al.]{O.~Petruk$^{1,2}$, S.~Orlando$^{3}$, V.~Beshley$^{2}$, F.~Bocchino$^{3}$\\
$^{1}$Institute for Applied Problems in Mechanics and Mathematics, Naukova St.\ 3-b,
   79060 Lviv, Ukraine\\
$^{2}$Astronomical Observatory, National University, Kyryla and Methodia St.\ 8, 79008 Lviv, Ukraine\\
$^{3}$INAF - Osservatorio Astronomico di Palermo ``G.S.
              Vaiana'', Piazza del Parlamento 1, 90134 Palermo, Italy
}

\begin{document}

\date{Accepted .... Received ...; in original form ...}

\pagerange{\pageref{firstpage}--\pageref{lastpage}} \pubyear{2010}

\maketitle

\label{firstpage}

\begin{abstract}
Distributions of nonthermal surface brightness of supernova remnants
(SNRs) contain important information about the properties of magnetic
field and acceleration of charged particles. In the present paper, the
synchrotron radio, X-ray, and inverse-Compton (IC) \g-ray
maps of adiabatic SNRs in uniform interstellar medium and interstellar
magnetic field are modeled and their morphology is analyzed, with
particular emphasis to comparison of azimuthal and radial variations of
brightness in radio, X-rays, and \g-rays.  
Approximate analytical formulae for the azimuthal and radial profiles of
the synchrotron radio and X-ray as well as the IC \g-ray brightness
are derived. They reveal the main factors which influence the pattern of
the surface brightness distribution due to leptonic emission processes
in shells of SNRs and can account for some non-linear effects of
acceleration if necessary. These approximations provide observers and
theorists with a set of simple diagnostic tools for quick analysis of
the non-thermal maps of SNRs.
\end{abstract}


\section{Introduction}

Non-thermal images of SNRs are rich sources of information about the
properties of interstellar magnetic field (ISMF) behavior and kinetics
of charged particles in vicinity of the strong non-relativistic shocks.
Despite of their importance, images of SNRs -- in contrast to broad-band
spectra -- are not well studied.

Synchrotron X-ray brightness profiles were used as diagnostic
tools for the estimate of the post-shock magnetic field in some SNRs
\citep[e.g.][]{Ber-Volk-2004-mf}. Radio azimuthal profiles were used
for determination of some properties of SN~1006 \citep[][hereafter
Paper I]{pet-SN1006mf} and X-ray radial profiles were used to detect
the shock precursor in SN~1006 and prove particle acceleration
\citep{Morlino-etal-2010}.

A detailed approach to modeling the {synchrotron} images of adiabatic
SNRs in uniform ISMF and uniform interstellar medium (ISM) is developed
by \citet{Reyn-98}. \citet{reyn-fulbr-90,Reyn-98,Reyn-04} use modeled
synchrotron maps of SNRs to put constraints on properties of accelerated
particles. Properties of the inverse-Compton (IC) \g-ray maps are
investigated and compared to radio images in \citet[][hereafter Paper
II]{thetak}.

The influence of nonuniform ISM and/or nonuniform ISMF on
the thermal X-ray morphology of adiabatic SNRs are studied in
\citet{1999A&A...344..295H}, on the radio maps in \citet{Orletal07} and
on the synchrotron X-ray and IC \g-ray images in {\citet{Orletal10}}.

All studies of SNR maps assume classic MHD and test-particle theory of
acceleration. Though they neglect effects of the back-reaction of the
efficiently accelerated particles, they are able to explain general
properties of the distribution of the surface brightness in radio,
X-rays and \g-rays. This is because the classic theory, in contrast to
the non-linear one, is able to deal with oblique shocks, that is vital
for synthesis of SNR images.

At present, the theory which considers effects of accelerated
particles on the shock and on acceleration itself is developed for
the initially quasi-parallel shocks only. One may therefore model
the only radial profiles of brightness, in the rather narrow region
close to the shock (in order to be certain that obliquity does not
introduce prominent modifications). Effects of non-linear
acceleration on the radial profiles of brightness are considered in
\citet{Ell-Cassam2005-profiles,Decours-2005-prof,Ell2008-images,Zirakashv-Aha-2010}.

Future studies on SNR morphology should take into account the NLA
effects. Nevertheless, the classic approach is able to reveal the
general properties of SNR maps determined by MF behavior and particle
acceleration.  Beside that, it is important to know the properties of the
`classic' images because they create the reference base for investigation
of the efficiency of NLA effects in the surface brightness distribution
of SNRs.

In Paper I, we introduced a method to derive an aspect angle of ISMF
from the radio brightness of SNR.  In Paper II, we synthesized radio and
IC \g-ray maps and concluded that {coincidence of the position of the \g-ray
and radio limbs is not a common case in theoretical models, because 
different parameters are dominant in determintion of the radio and \g-ray brightness variations.
On the other hand, radio, X-ray and \g-ray observations \citep{pet-SN1006mf,SN1006Marco,HESS-sn1006-2010} show
that radio, X-ray and $\gamma$-ray limbs coincide in SN 1006. As discussed
in Paper II, such coincidence might be attributed to a combination of
obliquity dependences of magnetic field and properties of emitting
particles, as well as orientation versus the observer.} 

In the present
paper, we make a step forward extending our model of non-thermal
leptonic emission of Sedov SNRs in uniform medium to the X-ray band.
We also derive brightness profiles for representative parameters which
are suited for the comparison with adiabatic SNRs.  Moreover, we derive
analytical approximations of the azimuthal and radial profiles of 
radio, X-ray and \g-ray brightness which are extremely useful to 
demonstrate their dependence on
the acceleration parameters, magnetic field orientation and the viewing
geometry; they can also be directly and very easily fitted to SNR images
to derive estimations for the best-fit quantities. While the analytical 
approximation cannot
substitute the more accurate numerical simulations, we show that they
retain enough accuracy to represent an effective diagnostic tool for
the study of non-thermal SNR shells.

\section{Rigorous 
treatment of synchrotron and IC emission
of Sedov SNRs} \label{xmaps:xray_maps}

\subsection{General considerations}

{The present paper is continuation of the study in Paper II. 
The model and numerical realisation are therefore similar to those 
used in the Paper II; 
the reader is referred to this paper for more details. In short,}
in order to investigate the properties of the leptonic emission of
Sedov SNR, we use the \citet{Sedov-59} solutions for dynamics of
fluid as well as the \citet{Reyn-98} description of the MF behavior
downstream of the shock. The use of analytical solutions allows us to
reduce the computational time considerably. We follow \citet{Reyn-98}
also in calculation of the evolution of the distribution function
$N(E)$ of relativistic electrons downstream of the shock (see Appendix
\ref{xmaps:app2} for details).

The ISMF orientation versus observer is described by the aspect angle
$\phi\rs{o}$, an angle between ISMF and the line of sight.
Let us define also the obliquity angle $\Theta\rs{o}$ as the angle between
ISMF and the normal to the shock, and the azimuth angle $\varphi$ in the
projection plane which is measured from the direction of the component
of ISMF in the plane of the sky (see Fig.~1 in Paper II).

The compression factor for ISMF at the shock front $\sigma\rs{B}$ is
modulated from unity at parallel shock to $\sigma=(\gamma+1)/(\gamma-1)$
at perpendicular one (where $\gamma$ is the adiabatic index), in agreement
with \citet{Reyn-98}.

At the shock, the energy spectrum of relativistic electrons is taken to
be $N(E) = KE^{-s}\exp\left(-(E/E\rs{max})^\alpha\right)$, where
$E\rs{max}$ is the maximum energy of electrons, $s$ the spectral index,
$K$ the normalization, {free parameter $\alpha$ regulates the rate of the spectrum decrease around 
the high-energy end\footnote{{A number of observations \citep{Ell-et-al-2000,Ell-et-al-2001,Lazendic-et-al-2004,Uchiyama-et-al-2003}
reveal evidence of broadening of 
the high-energy cut-off, i.e. $\alpha<1$. Such broadening should be attributed to the physics of acceleration \citep{Pet06}, 
rather than to the effect of superposition of spectra in different conditions 
along the line of sight as it was suggested initially by \citet{Reyn96}.
From other side, theoretical model of \citet{Zirakashv-Aha-2007} demonstrates that 
it should be $\alpha=2$ for the loss-limited models, whereas in practice,
taking the time evolution into account it could be between 1 and 2 \citep{Schure-et-al-2010}. 
Also \citet{Kang-Ryu-2010} suggest that $\alpha>1$.}}.} 
Evolution of the `electron
injection ability' of the shock is represented as $K\propto V^{-b}$ where
$V$ is the shock velocity, and $b$ is a constant. The variation of the
distribution $N(E)$ over the surface of the SNR and its evolution downstream
of the shock are calculated as described by \citet{Reyn-98}. 
{Following \citet{Reyn-98}, we synthesize the synchrotron and IC emission,
considering each of the three cases of variation of electron injection
efficiency with the shock obliquity (quasi-perpendicular, isotropic, and
quasi-parallel particle injection) and each of the three alternatives for
time and spatial dependence of $E\rs{max}$ (time-limited, loss-limited
and escape-limited with a gyrofactor $\eta$ as a parameter; it is a
ratio of the mean free path of a particle to its gyroradius).}
Both the synchrotron and the IC losses are included as channels for
the radiative losses of relativistic electrons, though the IC losses
are quite small comparing to synchrotron due to large MF considered.

The surface brightness is calculated by integrating emissivities
along the line of sight within the SNR. The emissivity of electrons is
given by
\begin{equation}
 q(\varepsilon)=\int_{0}^{\infty}N(E)p(E,\varepsilon)dE  
 \label{IC-emiss}
\end{equation}
where $\varepsilon$ is the photon energy and $p$ is the spectral
distribution of synchrotron or IC radiation power of electron. 
We assume that information about orientation of $B$ inside SNR is lost
because of turbulence, in practice, we use an average aspect angle
downstream.
{The distribution $p$ is calculated with the use of the analytical 
approximation developed by \citet[][see also Paper II]{Pet08IC}.}

\begin{figure*}
  \centering
  \includegraphics[width=16.cm]{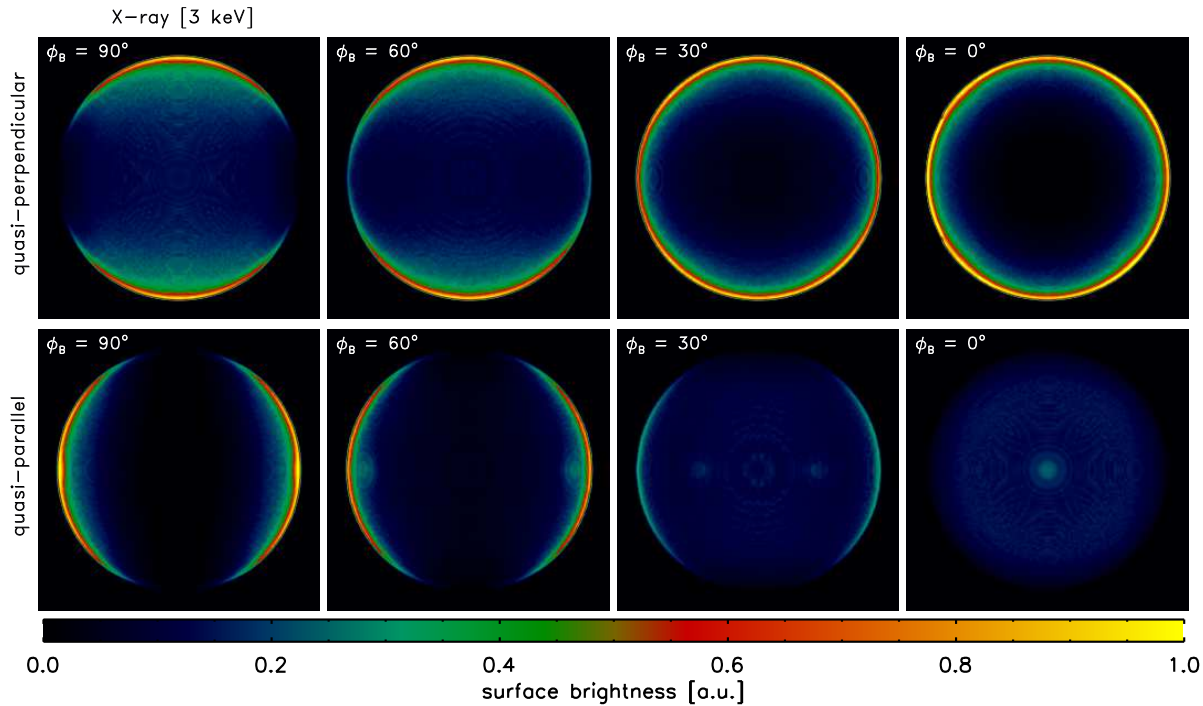}
  \caption{Maps of synchrotron X-ray
           surface brightness synthesized for different
           aspect angles (labeled in the upper left corner of each
           panel). The figure shows the quasi-perpendicular (top), and
           quasi-parallel (bottom) particle injection models. The model
           of $E\rs{max}$ is time-limited with $\eta=1$.
           Calculations are done for $E\rs{max\|}=26\un{TeV}$, 
           $B\rs{o}=30\un{\mu G}$, $s=2$, $\alpha=0.5$, $b=-3/2$. 
           The ambient magnetic field is along the horizontal axis. }
  \label{xmaps:fig3}
\end{figure*}

\subsection{Images}

The resulting synchrotron radio images and the IC \g-ray
images synthesized by our model have been already presented in 
Paper I and Paper II. Therefore, we present 
here only the X-ray images {(see Fig.~\ref{xmaps:fig3})}, adding appropriate references to the previous
work to let the reader do a quick comparison.

The pattern of synchrotron X-ray brightness of SNR is in general similar
to the radio one.  In most cases, the bright X-ray limbs or other features
are located in the same azimuth as in the radio images.  The only
differences appear due to radiative losses which modify downstream
distribution of the electrons emitting in X-rays (thus the features of
brightness are radially thinner) and due to variation of $E\rs{max}$
over the SNR surface.  In the radio (see {Fig.~4} in Paper II) as also
in the X-ray band, the remnant shows two symmetric bright lobes (for
$\phi\rs{o} = 90^\mathrm{o}$) in all the injection models with the maxima
in surface brightness coincident in the two bands. The maxima are located
at perpendicular shocks in the quasi-perpendicular and isotropic models
(i.e. where $B$ is higher), and at parallel shocks in the quasi-parallel
model (i.e. where emitting electrons are only presented).  The lobes
are much radially thinner in X-rays than in radio because of the large
radiative losses at the highest energies that make the X-ray emission
dominated by radii closest to the shock.

The X-ray morphology of SNR is different for different aspect angles
(Fig.~\ref{xmaps:fig3}, cf. with {Figs.~4, 5 and 6 of Paper II for} radio and \g-ray images, {respectively}).
In the case of quasi-perpendicular injection, the morphology is bilateral
(two lobes) for large aspect angles ($\phi\rs{o} > 60^0$, i.e. the
component of ISMF in the plane of the sky is larger than that along the
line of sight) and almost ring-like for low aspect angles ($\phi\rs{o}
< 30^0$; see Fig.~\ref{xmaps:fig3}) with intermediate morphology
between $30^0$ and $60^0$. In the case of quasi-parallel injection,
the remnant morphology in the radio band is known to be bilateral
for large aspect angles and characterized by one or two eyes for low
aspect angles \citep[][]{reyn-fulbr-90,Orletal07}. On the other hand,
it is worth noting that the remnant morphology in X-rays is in general
bilateral for aspect angles $\phi > 30^0$ and centrally bright for very
low angles, indeed a rather limited set of possible cases (lower panels
in Fig.~\ref{xmaps:fig3}). This happens because the non-thermal X-ray
emission originates from a very thin shell behind the shock making the
effect of limb brightening in X-rays more important than in the radio
band. In addition, we note that centrally bright X-ray (and radio)
SNRs are expected to be much fainter than bilateral SNRs (see lower
panels in Fig.~\ref{xmaps:fig3}) and consequently much more difficult
to be observed. The above considerations may affect the statistical
arguments generally invoked against the quasi-parallel injection
(i.e. the fact that this model produces morphology which is not observed;
e.g. \citealt{reyn-fulbr-90, Orletal07}).


Images on Fig.~\ref{xmaps:fig3} are calculated 
{assuming a dependency of $E\rs{max}$
on the obliquity angle which corresponds to the time-limited model with
$\eta = 1$ as introduced by \citet{Reyn-98}, namely $E\rs{max} \propto
{\cal E}\rs{max}(\Theta\rs{o})B\rs{o}$, where ${\cal E}\rs{max}(\Theta\rs{o})$
is a function describing smooth variations of $E\rs{max}$
versus obliquity, and $B\rs{o}$ is the pre-shock ISMF strength
\citep[in such a model no dependency on the shock velocity is present;][]{Reyn-98}.
With this particular choice,} 
$E\rs{max}$ is quite similar for
different obliquities, namely $E\rs{max\bot}/E\rs{max\|}=1.3$. Larger
values of $\eta$ always provide $E\rs{max\bot}/E\rs{max\|}>1$ in this model,  
thus the character of azimuthal variation of brightness would be similar. 


The efficiency of variation of $E\rs{max}$ with obliquity in
modification of the azimuthal distribution of X-ray synchrotron
brightness depends obviously on the photon energy: if the maximum
contribution to the emission at given photon energy is from
electrons with energy much less than $E\rs{max}$ then this effect
is negligible.  It is useful to introduce the reduced photon energy,
as $\tilde \varepsilon=\nu/\nu\rs{c}(E\rs{max\|},B\rs{o})$ where
$\nu\rs{c}(E,B)=c_1\left\langle\sin\phi\right\rangle E^2B$ is the
synchrotron characteristic frequency, $c_1=6.26\E{18}\un{cgs}$, or
\begin{equation}
 \tilde \varepsilon=19\ \varepsilon\rs{keV}
                \left(\frac{E\rs{max\|}}{10\un{TeV}}\right)^{-2}
                \left(\frac{B\rs{o}}{10\un{\mu G}}\right)^{-1},
\label{xmaps:tilde_e_def}
\end{equation}
where $\varepsilon\rs{keV}$ is the photon energy in keV.
If $\tilde\varepsilon=0.29$ then most of the contribution 
to the synchrotron X-ray emission is from electrons with energy
$E=E\rs{max}$.\footnote{For reference: the maximum contribution to
synchrotron X-ray emission at 3 keV in MF 30 $\un{\mu G}$ is from
electrons with energies 72 TeV; the maximum contribution to IC \g-ray
emission at 1 TeV originates from electrons with energies 17 TeV.}
Figs.~\ref{xmaps:fig3} is 
calculated for
$\tilde \varepsilon=2.8$, i.e. images shown are mainly due to emission of
electrons with energy few times higher than $E\rs{max\|}$;
the role of variation of $E\rs{max}$ is therefore clearly visible
in the images.

\begin{figure}
 \centering
 \includegraphics[width=8truecm]{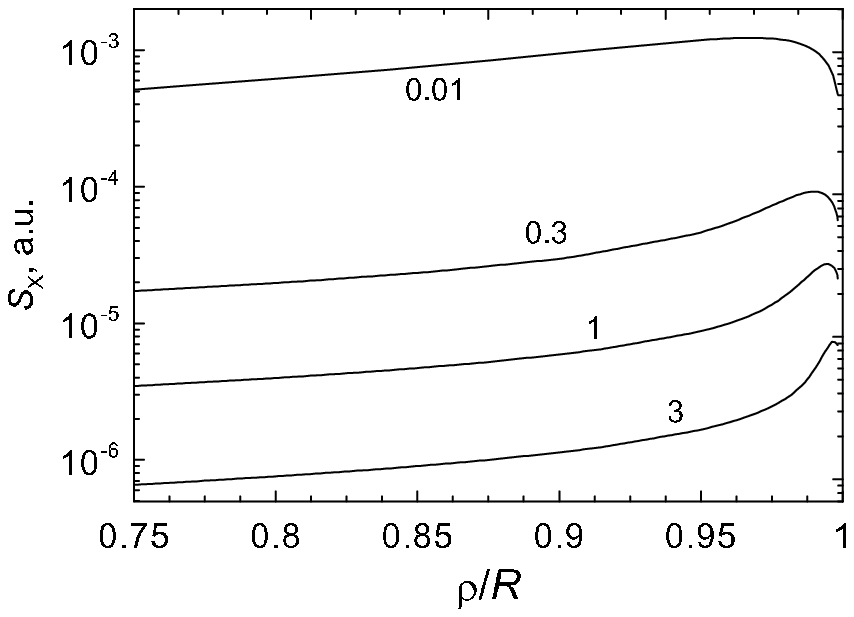}
 \caption{Radial profiles of X-ray surface brightness for different photon energies  
 $\tilde\varepsilon$ (marked near respective lines). 
 Calculations are done for $\phi\rs{o}=0$, $\varphi=0$, $b=-3/2$, $\epsilon\rs{f\|}=1$. 
                }
 \label{xmaps:fig_photon_energy}
\end{figure}
\begin{figure}
 \centering
 \includegraphics[width=7.7truecm]{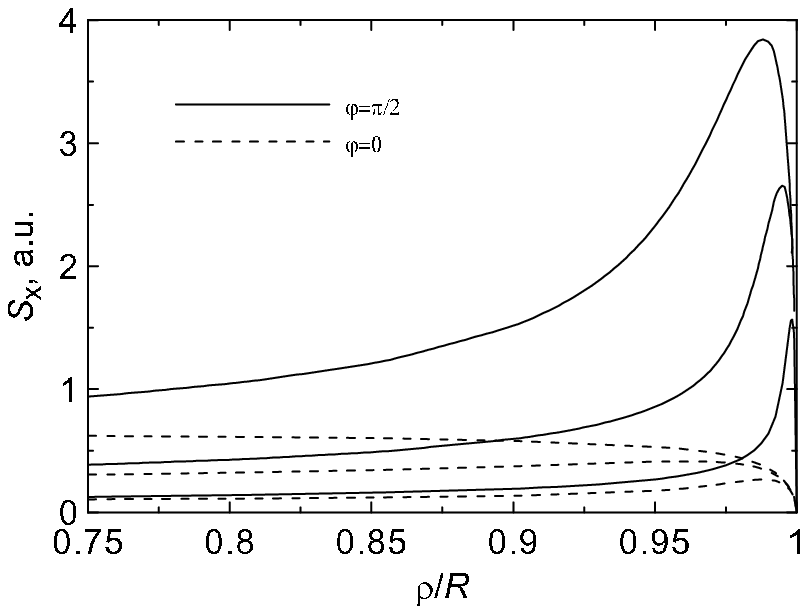}
 \caption{Effect of radiative losses (represented by $\epsilon\rs{f\|}$) on the radial profiles of X-ray surface brightness, at two azimuth. 
 The reduced fiducial energy $\epsilon\rs{f\|}=0.3,1,3$ (from below); 
 $\phi\rs{o}=\pi/2$, $\tilde\varepsilon=1$, $b=0$. 
                }
 \label{xmaps:fig_fiducial}
\end{figure}

\subsection{Brigthness profiles}

In the present subsection, $E\rs{max}$ is assumed to be constant in time
and the same for any obliquity; in addition, isotropic injection, $s=2$
and $\alpha=1$ in the energy spectrum of relativistic electrons
are assumed.

The radial thickness of features in the X-ray images is sensitive
to the photon energy: the larger the energy the thinner the limbs
(Fig.~\ref{xmaps:fig_photon_energy}). This is because radiative losses
$\dot E$ of electrons with energy $E$ is efficient for more energetic
electrons, $\dot E\propto E^2$. If $\tilde\varepsilon>0.29$ then 
most of the contribution to the synchrotron X-ray emission is from
electrons with energies $E>E\rs{max}$ where the radiative losses are of
the main importance.

An important factor for emission of highly energetic electrons is the
fiducial energy, which reflects the importance of radiative losses in
modification of the electron distribution. It is defined
as $\epsilon\rs{f}=637\left(B\rs{s}^2 t E\rs{max}\right)^{-1}$ \citep{Reyn-98}, or 
\begin{equation}
 \epsilon\rs{f}=13 
 \left(\frac{B\rs{s}}{10\un{\mu G}}\right)^{-2}
 \left(\frac{E\rs{max}}{10\un{TeV}}\right)^{-1}
 \left(\frac{t}{1000\un{yrs}}\right)^{-1}.
 \label{xmaps:etaf}
\end{equation}
Radiative losses are important for $\epsilon\rs{f}<1$ and minor for $\epsilon\rs{f}>1$. 
Fig.~\ref{xmaps:fig_fiducial} demonstrates how the value of $\epsilon\rs{f}$ affects the radial profiles of X-ray 
brightness: the smaller $\epsilon\rs{f}$ the thinner the rim.

\begin{figure}
 \centering
 \includegraphics[width=8truecm]{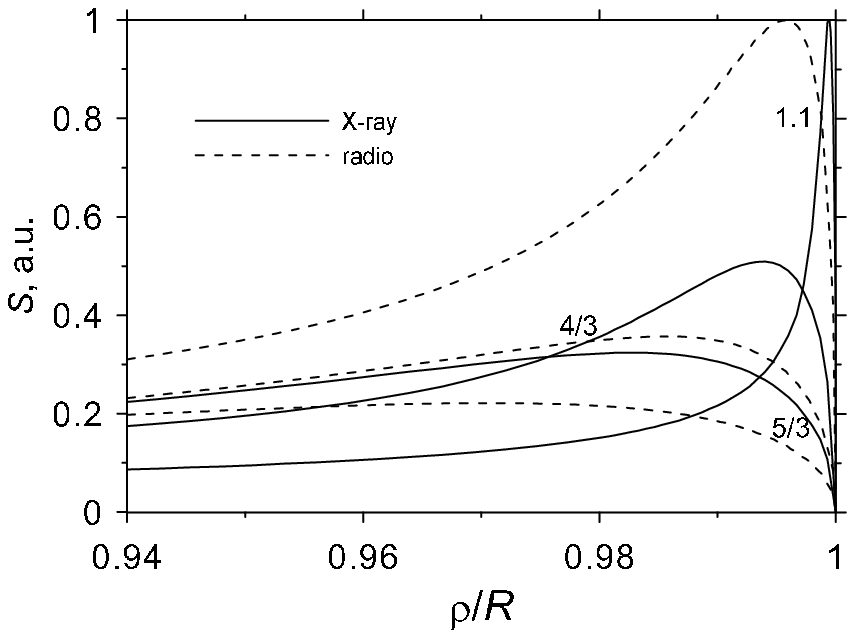}
 \caption{Radio (dashed lines) and X-ray (solid lines) radial profiles of surface brightness 
 for the adiabatic index $\gamma=5/3$, $4/3$ and $1.1$. 
 Calculations are done for $\phi\rs{o}=0$, $\varphi=0$, $b=0$, 
 $\epsilon\rs{f\|}=3$, $\tilde\varepsilon=0.3$. 
                }
 \label{xmaps:fig_gamma}
\end{figure}

Our model does not include consistently the effects on shock dynamics
due to back-reaction of accelerated CRs. However, we may approach the
effect of shock modification by considering different values of the
adiabatic index $\gamma$ which is expected to drop from the value of
an ideal monoatomic gas. In particular, Fig.~\ref{xmaps:fig_gamma}
considers the cases of $\gamma=5/3$ (for an ideal monatomic gas),
$\gamma = 4/3$ (for a gas dominated by relativistic particles), and
$\gamma = 1.1$ (for large energy drain from the shock region due to
the escape of high energy CRs). The shock modification results in more
compressible plasma and, therefore, in the radially-thinner features of
the nonthermal images of SNRs.  A small distance between the forward
shock and contact discontinuity \citep{chr08,SN1006Marco} could also
be attributed to $\gamma<5/3$.  Effect of the index $\gamma$ on the
profiles of hydrodynamical parameters downstream of the adiabatic shock
is widely studied \citep[e.g.][]{Sedov-59}: smaller $\gamma$ makes the
shock compression factor higher,
\begin{equation}
 \sigma=\left\{
 \begin{array}{ll}
 	4&\mathrm{for}\ \gamma=5/3,\\
	7&\mathrm{for}\ \gamma=4/3,\\
	21&\mathrm{for}\ \gamma=1.1,\\
 \end{array}
 \right.
\end{equation} 
and the gradient of density downstream stronger (e.g. Appendix~\ref{xmaps:app3}),
\begin{equation}
 \bar n(\bar r)\approx \bar r^{\kappa\rs{nr}},\qquad 
 \kappa\rs{nr}=\left\{
 \begin{array}{ll}
 	12&\mathrm{for}\ \gamma=5/3,\\
	25&\mathrm{for}\ \gamma=4/3,\\
	88&\mathrm{for}\ \gamma=1.1,\\
 \end{array}
 \right.
\end{equation} 
({where $\bar n=n/n\rs{s}$, $n\rs{s}$ is the density immediately postshock and} $\bar r=r/R \leq 1$).
In addition to that, the X-ray (and also TeV \g-ray) brightness is modified by increased radiative 
losses of emitting electrons. Really, the larger compression leads to the higher post-shock MF and thus to increased 
losses, $\dot E\propto B^2$, which results in turn in the thinner radial profiles of brightness.  

\begin{figure*}
 \centering
 \includegraphics[angle=270,width=17.6truecm]{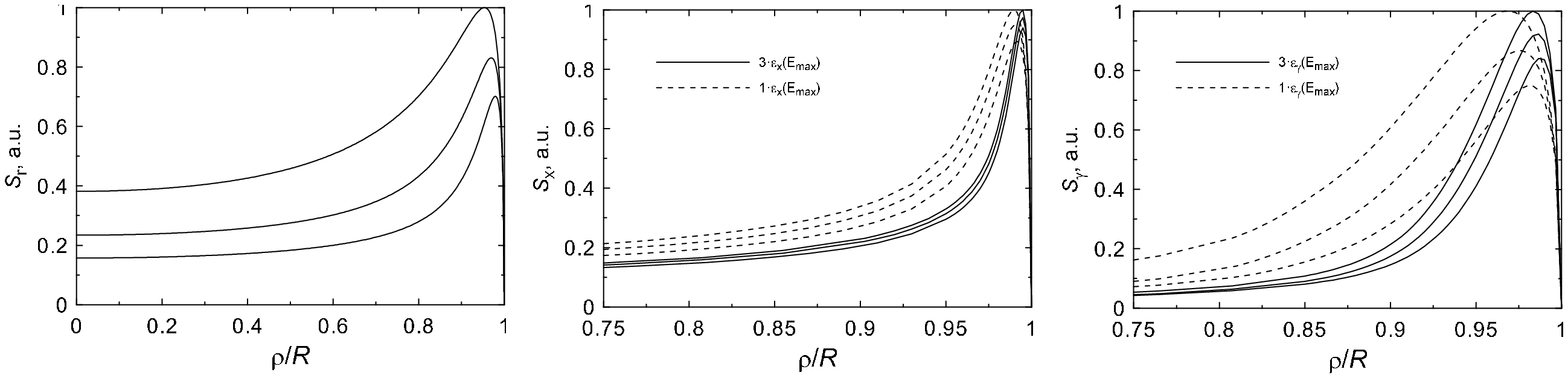}
 \caption{Evolution of injection efficiency (represented by $b$) and radial profiles of surface brightness 
 in radio (left), X-rays (middle) and \g-rays (right). 
 The parameter in relation $K\rs{s}\propto V^{-b}$ is $b=-3/2,0,2$ {(from above)}. 
 Other parameters are $\phi\rs{o}=\pi/2$, 
 $\varphi=0$ (for radio and X-rays) and $\varphi=\pi/2$ (for \g-rays), $\epsilon\rs{f\|}=1$. 
 X-ray and \g-ray profiles are shown for two photon energies: where the most contribution is from electrons with 
 $E=E\rs{max}$ (dashed lines) and three times larger (solid lines); 
 in case of X-rays, solid lines correspond to $\tilde\varepsilon=1$. 
                }
 \label{xmaps:fig_b}
\end{figure*}
\begin{figure*}
 \centering
 \includegraphics[angle=270,width=17.6truecm]{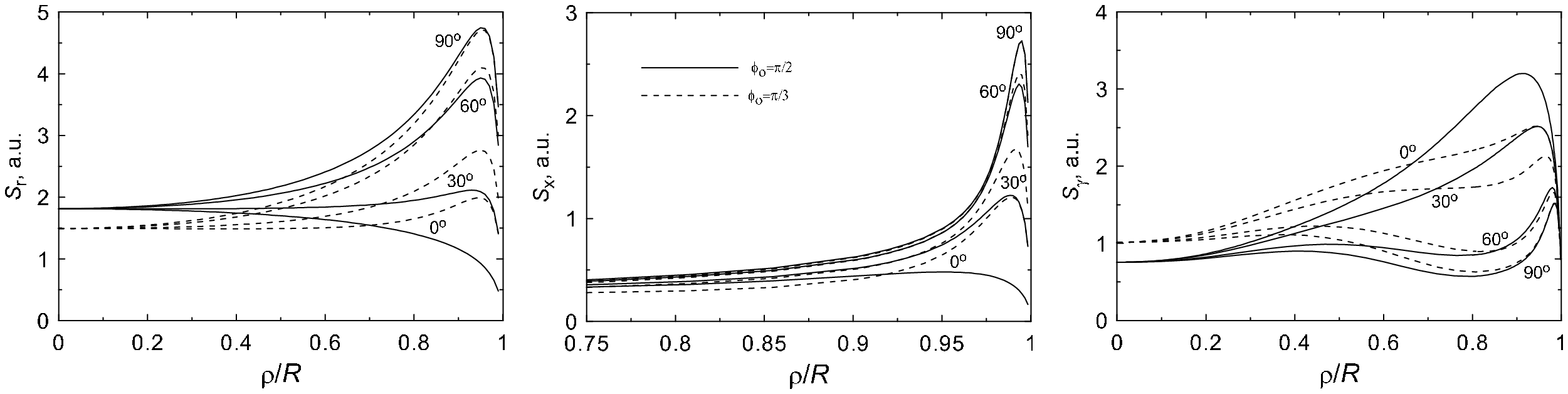}
 \caption{Radial profiles of the radio (left), X-ray (middle) and \g-ray (right) surface brightness of SNR for 
 different azimuth: $0,\pi/6,\pi/3,\pi/2$ (marked near lines). 
 The aspect angles are $\pi/2$ (solid lines) and $\pi/3$ (dashed lines); 
 $b=-3/2$, $\epsilon\rs{f\|}=1$, $\tilde\varepsilon=1$, \g-ray profiles are for photons with energy $0.1\varepsilon\rs{\gamma}(E\rs{max})$. 
                }
 \label{xmaps:fig_azimuth}
\end{figure*}

It is unknown how the injection efficiency (the fraction of nonthermal
particles) depends on the properties of the shock. We parameterized its
evolution as $K\rs{s}\propto V^{-b}$ where $b$ is a constant. Effect of
the parameter $b$ on the radial profiles of the surface brightness is
demonstrated on Fig.~\ref{xmaps:fig_b}. The smaller $b$ the thicker
the profiles, because there are more emitting electrons in deeper
layers, which were injected at previous times. This property affects
the nonthermal emission in all bands. However, the effect is less
prominent in X-rays (and in TeV \g-rays) if radiative losses are quite
effective to dominate it (see Fig.~\ref{xmaps:fig_b}, lines for different
photon energies). 
{This is in agreement with finding of \citet{Parizot-et-al-2006} and \citet{Vink-et-al-2006} who showed that, 
for $E>E\rs{max}$, the width of the synchrotron limbs does not 
depend on the shock velocity in the loss limited case.
Instead,} profiles of the radio brightness may be used to put
limitations on the value of $b$.

In a similar fashion, the X-ray and \g-ray radial profiles are affected
also by the time evolution of the maximum energy, $E\rs{max}\propto
V^{q}$. However, it seems impossible to determine $q$ from such profiles
because contribution of other factors is often dominant.

An interesting feature of the synchrotron images of SNRs is apparent
from Fig.~\ref{xmaps:fig_azimuth}. The maxima of the radial profiles of
brightness for different azimuth are located almost at the same distance
$\rho$ from the center of projection, for radio and X-rays.
Thus, the best way to analyze the azimuthal profiles of the surface
brightness is to find the position $\rho$ of the maximum for one
azimuth and then to trace the azimuthal profile of brightness
$S\rs{\rho}(\varphi)$ for fixed $\rho$.

\section{Explicit approximate analytical formulae for surface brightness profiles of non-thermal SNR shells}
\label{xmaps:approximation_section}

The rigorous model discussed in the previous section is able
to predict the non-thermal emission of Sedov SNR under a variety
of conditions. However, in practical applications, it may be rather
time-consuming to perform an extensive parameter space exploration, and
the crucial dependence on the relevant parameters of the acceleration
processes may be hidden.

In order to understand how the properties of MF, electron injection and
acceleration influence the brightness distribution, we derived analytic
approximate formulae for the azimuthal and radial profiles of the surface
brightness of adiabatic SNR. In this way, we can easily see what are the
main factors which determine the pattern of the nonthermal images of SNRs,
and which of them are mostly responsible for the azimuthal variation of
the surface brightness and which for the radial one. This turns to be
extremely useful in guiding the comparison with real observations. The
analytical formulae are valid close to the shock only, but are adequate
to describe azimuthal and radial variations of brightness around maxima
which are located close to the edge of SNR shells.

\subsection{Radio profiles}

Let the evolution and obliquity variation of the electron injection
efficiency be denoted as $V(t)^{-b}\varsigma(\Theta\rs{o})$ and of the
obliquity variation of MF compression{/amplification} as $\sigma\rs{B}(\Theta\rs{o})$; 
{for the sake of generality we assume $\varsigma(\Theta\rs{o})$ and 
$\sigma\rs{B}(\Theta\rs{o})$ to be some arbitrary smooth functions}. 
Properties of the azimuthal and radial profiles of the radio
brightness is determined mostly by (Appendix \ref{xmaps:app5})
\begin{equation}
 S\rs{r}(\varphi,\bar\varrho)\propto
 \varsigma\big(\Theta\rs{o,eff}\left(\varphi,\phi\rs{o}\right)\big)
 \sigma\rs{B}\big(\Theta\rs{o,eff}\left(\varphi,\phi\rs{o}\right)\big)^{(s+1)/2}
 I\rs{r}(\bar\varrho)
 \label{xmaps:radio_brightness}
\end{equation}
where 
\begin{equation}
 I\rs{r}=
 {1\over \sqrt{1-\bar \varrho^2}} 
 {1-\bar\varrho^{\sigma(\kappa\rs{r}+1)}\over \sigma(\kappa\rs{r}+1)},
 \label{xmaps:Iradio}
\end{equation}
$\bar\varrho=\rho/R$, $\sigma$ is the shock compression ratio, 
\begin{equation}
 \kappa\rs{r}={3b}/{2}+(2+s)\kappa\rs{ad}+{1}/{\sigma}+s,
\end{equation}
$\kappa\rs{ad}$ is close to unity for $\gamma=1.1\div5/3$ (Appendix \ref{xmaps:app3}). 
The effective obliquity angle $\Theta\rs{o,eff}$ is related to azimuth $\varphi$ and aspect $\phi\rs{o}$ as
\begin{equation}
 \cos\Theta\rs{o,eff}\left(\varphi,\phi\rs{o}\right)=\cos\varphi\sin\phi\rs{o},
\end{equation}
the azimuth angle is measured from the direction of ISMF in the plane of the sky.
Eq.~(\ref{xmaps:radio_brightness}) is a generalisation of the approximate formula derived in Paper I.

Eq.~(\ref{xmaps:radio_brightness}) shows that the azimuthal variation of the 
radio surface brightness $S\rs{\varrho}(\varphi)$ at a fixed radius $\varrho$ of projection, 
is mostly determined by the variations of the magnetic field compression 
(and amplification, if any)
$\sigma\rs{B}$ and by the variation of the electron injection efficiency $\varsigma$. 
The radial profile $S\rs{\varphi}(\varrho)\propto I\rs{r}(\varrho)$ is determined mostly by $\sigma$, $b$ and $s$. 
Adiabatic index $\gamma$ affects the radial and azimuthal profiles mostly through the compression factor 
$\sigma=(\gamma+1)/(\gamma-1)$ because $\kappa\rs{ad}$ weakly depends on $\gamma$. 

\subsection{Synchrotron X-ray profiles}

Let us assume that the maximum energy is expressed as $E\rs{max}(\Theta\rs{o},t)\propto V(t)^q {\cal E}\rs{max}(\Theta\rs{o})$, 
{where $q$ is a constant and ${\cal E}\rs{max}(\Theta\rs{o})$, for the sake of generality, is some arbitrary function describing the smooth variation of $E\rs{max}$ versus obliquity.}
The synchrotron X-ray brightness close to the forward shock is approximately (Appendix \ref{xmaps:app1})
\begin{equation}
 S\rs{x}(\varphi,\bar\varrho)\propto
 \varsigma(\varphi)\sigma\rs{B}(\varphi)^{(s+1)/2}
 \exp\left[-\left(\frac{\epsilon\rs{m}(\varphi)}
 {{\cal E}\rs{max}(\varphi)}\right)^{\alpha}\right]
 I\rs{rx}(\varphi,\bar\varrho)
 \label{xmaps:xray_brightness}
\end{equation}
where 
\begin{equation}
 I\rs{rx}=I\rs{r}(\bar\varrho)I\rs{x}(\varphi,\bar\varrho) 
  \label{xmaps:Ixray_main}
\end{equation} 
with  
\begin{equation}
 I\rs{x}=
 \left[1-\frac{\epsilon\rs{m}^{\alpha}(\psi+1)\alpha}{{\cal E}\rs{max}^{\alpha}}
 \left(1-\frac{1-\bar\varrho^{\sigma(\kappa\rs{r}+2)}}{1-\bar\varrho^{\sigma(\kappa\rs{r}+1)}}
 \frac{\kappa\rs{r}+1}{\kappa\rs{r}+2}\right)
 \right].
 \label{xmaps:Ixray}
\end{equation}
The parameter
\begin{equation}
 \psi=\kappa\rs{ad}+\frac{5\sigma\rs{B}^{2}\epsilon\rs{m}}{2\epsilon\rs{f\|}}-\frac{3q}{2}
 \label{xmaps:psi}
\end{equation}
is responsible for the adiabatic (the first term) and radiative (the second term) losses of emitting electrons and the time evolution of $E\rs{max}$ on the shock (the third term). 
The reduced electron energy which gives the maximum contribution to emission of photons with energy $\tilde\varepsilon$ is 
\begin{equation}
 \epsilon\rs{m}=\left(\frac{\tilde\varepsilon}{0.29\sigma\rs{B}}\right)^{1/2},
 \label{xmaps:epmX}
\end{equation}
it varies with obliquity (since MF varies; electrons with different energies contribute to the synchrotron emission at $\tilde\varepsilon$).
Parameters $\psi$, $\epsilon\rs{m}$, ${\cal E}\rs{max}$ depend on $\Theta\rs{o,eff}$ and, therefore, on the aspect angle $\phi\rs{o}$ and the azimuth angle $\varphi$.

If $\epsilon\rs{m}\ll 1$ then Eq.~(\ref{xmaps:xray_brightness}) for the X-ray brightness transforms to Eq.~(\ref{xmaps:radio_brightness}) for the radio brightness.

The thickness of the hard X-ray radial profile is used to estimate the post-shock strength of MF in a number of SNRs \citep[e.g.][]{Ber-Volk-2004-mf}. 
The absolute value of MF is present in Eq.~(\ref{xmaps:xray_brightness}) through $\tilde\varepsilon$ and $\epsilon\rs{f}$, Eqs.~(\ref{xmaps:tilde_e_def}), (\ref{xmaps:etaf}) which appear in $\psi$ and $\epsilon\rs{m}$, Eqs.~(\ref{xmaps:psi}), (\ref{xmaps:epmX}). In both cases, $B\rs{s}$ is in combination with $E\rs{max}$ (thus, the value of the electron maximum energy may affect the estimations). 
The idea of the method bases on the increased role of losses in X-rays due to larger MF, i.e. on the role of the second term in $\psi$, Eq.~(\ref{xmaps:psi}).
Really, the influence of $\kappa\rs{r}$ (i.e. of $s$ and $b$) is minor in X-rays (middle panel on Fig.~\ref{xmaps:fig_b}), if radiative losses affect the electron evolution downstream of the shock (i.e. for $\tilde\varepsilon\gsim 0.29$, $\epsilon\rs{f}\lsim 1$). 
The role of the first and the third terms in $\psi$ are also minor in most cases 
($q=0$ for the time-limited and escape-limited models and unity for the loss-limited one) because the second term  $\gsim 10$. 
However, the adiabatic index makes an important effect on the thickness of the profile, mostly through $\sigma$ which appears in $\sigma\rs{B}$ and in $I\rs{x}$. 
Being smaller than $5/3$ (that is reasonable especially in the case of efficient acceleration, which is actually believed to be responsible for the large MF), the index may compete to some extend the role of losses, used in the method for estimation of MF (see e.g. Fig.~\ref{xmaps:fig_gamma}) that might lead to smaller estimates of MF strength. 

\subsection{IC gamma-ray profiles}

The IC \g-ray brightness may approximately be described as (Appendix \ref{xmaps:app4})
\begin{equation}
 S\rs{ic}(\varphi,\bar\varrho)\propto
 \varsigma(\varphi)
 \exp\left[-\left(\frac{\epsilon\rs{m}}{{\cal E}\rs{max}(\varphi)}\right)^{\alpha}\right]
 I\rs{ic}(\varphi,\bar\varrho)
 \label{xmaps:IC_brightness}
\end{equation}
where 
\begin{equation}
\begin{array}{l}
I\rs{ic}(\varphi,\bar\varrho)\approx
 \displaystyle
 {1\over \sqrt{1-\bar \varrho^2}} 
 {1-\bar\varrho^{\sigma(\kappa\rs{ic}+1)}\over \sigma(\kappa\rs{ic}+1)}
 \\ \\ \times\displaystyle
 \left[1-\frac{\epsilon\rs{m}^{\alpha}\psi\alpha}{{\cal E}\rs{max}^{\alpha}}
 \left(1-\frac{1-\bar\varrho^{\sigma(\kappa\rs{ic}+2)}}{1-\bar\varrho^{\sigma(\kappa\rs{ic}+1)}}
 \frac{\kappa\rs{ic}+1}{\kappa\rs{ic}+2}\right)
 \right],
 \label{xmaps:Iic_main}
\end{array}
\label{xmaps:Igtext}
\end{equation}
\begin{equation} 
 \kappa\rs{ic}={3b}/{2}+({2+s})\kappa\rs{ad}+{1}/{\sigma}-1.
\end{equation}
The expression for $\psi$ is the same as (\ref{xmaps:psi}) but $\epsilon\rs{m}$ is different:  
\begin{equation} 
 \epsilon\rs{m}
 =\frac{\varepsilon^{1/2}}{2(kT)^{1/2}\gamma\rs{max\|}},
\end{equation}
where $\varepsilon$ is the \g-ray photon energy, $T$ the temperature of the seed photon field, $\gamma\rs{max}$ the Lorentz factor of 
electrons with energy $E\rs{max}$, or 
\begin{equation} 
 \epsilon\rs{m}
 =1.66\left(\frac{E\rs{max\|}}{10\un{TeV}}\right)^{-1}\left(\frac{\varepsilon}{1\un{TeV}}\right)^{1/2}
 \left(\frac{T}{2.75\un{K}}\right)^{-1/2}.
\end{equation}
Eq.~(\ref{xmaps:IC_brightness}) is a generalisation of the approximate formula derived in Paper II.

The azimuthal variation of the IC \g-ray brightness depends mostly on the injection efficiency. The role of variation of $E\rs{max}$ is prominent only if obliquity dependence of injection is not strong. Parameter $\alpha$, being smaller than unity, results in smaller azimuthal contrasts of synchrotron X-ray or 
IC \g-ray brightness comparing to model with purely exponential cut-off in $N(E)$.
The radial distribution of IC brightness is determined mostly by $\sigma$, $\epsilon\rs{m}$, $\epsilon\rs{f}$, $b$ and, to the smaller extend, by $s$ and $q$.

\begin{figure*}
 \centering
 \includegraphics[clip,width=14truecm]{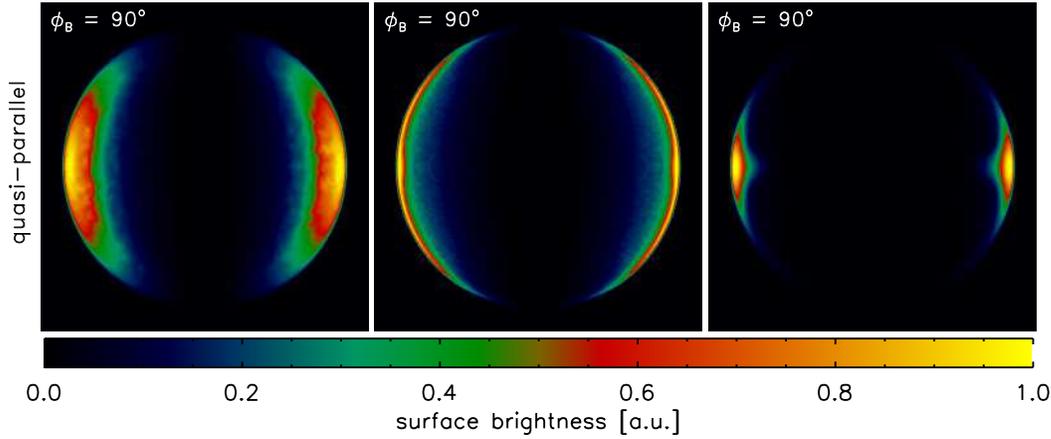}
 \caption{{Radio (left), X-ray (center) and \g-ray (right) images of the polar-cap SNR, assuming quasi-parallel injection.  
 Model parameters are the same as in Fig.~\ref{xmaps:fig3}. 
 Different azimuthal and radial sizes of limbs in various bands 
 are clearly visible.} 
                }
 \label{xmaps:limb_sizes}
\end{figure*}

\subsection{Accuracy of the formulae}

The approximations presented above do not require long and complicate numerical simulations but 
restore all the properties of nonthermal images discussed in the previous sections as well as in Papers I and II, including dependence on the aspect angle. 
{Therefore, they may be used as a simple diagnistic tool for non-thermal maps of SNRs.}

The formulae are rather accurate in description of the brightness distribution close to the shock. 
They do not represent centrally-brightened SNRs. Instead,  
they may be used in SNR shells for those azimuth $\varphi$ where $\epsilon\rs{m}\lsim 1$ and $\epsilon\rs{f}\gsim 0.1$, in the range of $\bar\varrho$ from $(1-2\Delta\bar\varrho\rs{m})$ to 1, where $\Delta\bar\varrho\rs{m}=1-\bar\varrho\rs{m}$, $\bar\varrho\rs{m}$ is the radius (close to the shock) where the maximum in the radial profile of brightness happens. 
Approximations are compared with numerical calculations in respective Appendices and their applicability 
is discussed in details on example of the IC emission in Appendix~\ref{xmaps:app4_accuracy}. 

\section{Discussion}

Analysis of azimuthal profiles of brightness in different bands allows
one to put limitations on models of injection, MF, $E\rs{max}$. In
most cases, the best way to estimate the azimuthal variation
$S\rs{\rho}(\varphi)$ of the surface brightness is the following. An
approximate radial profile $S\rs{\varphi}(\rho)$ of the brightness should
be produced for azimuth where the largest losses occur (i.e. where
$\epsilon\rs{f}\propto\left({\cal E}\rs{max}\sigma\rs{B}^2\right)^{-1}$ is
smaller; e.g. at $\varphi=90^\mathrm{o}$). This allows us to find
$\bar\varrho\rs{m}$ which should be used in $S\rs{\rho}(\varphi)$
in order to estimate the azimuthal variation of brightness. 
$\bar\varrho\rs{m}$ keeps us at maxima in the radial brightness profiles
for different azimuth (Fig.~\ref{xmaps:fig_azimuth}).

Energy of electrons evolve downstream of the shock as $E(\bar
a)=E\rs{i}{\cal E}\rs{ad}{\cal E}\rs{rad}$, where $E\rs{i}$ is initial
energy, $\bar a=a/R$, $a$ the Lagrangian coordinate. Adiabatic and
radiative losses of electrons in a given fluid element $a$ are represented
by functions ${\cal E}\rs{ad}(\bar a)\leq1$, ${\cal E}\rs{rad}(\bar
a)\leq1$ respectively (Appendix \ref{xmaps:app2}). Close to the shock,
they are approximately ${\cal E}\rs{ad}\approx \bar a^{\kappa\rs{ad}}$
($\kappa\rs{ad}$ depends on the adiabatic index $\gamma$ only and is
close to unity for $\gamma=1.1\div5/3$), ${\cal E}\rs{rad}\approx
\bar a^{{5\sigma\rs{B}^{2}\epsilon\rs{m}}/{(2\epsilon\rs{f\|})}}$
(Appendix \ref{xmaps:app3}).  The latter expression clearly shows that
the fiducial energy $\epsilon\rs{f}$ is important parameter reflecting
the `sensitivity' of the model to the radiative losses, as it is
shown by \citet{Reyn-98}: the larger the fiducial energy the smaller
the radiative losses.  In fact, ${\cal E}\rs{rad}=1$ means no radiative
losses at all. Another fact directly visible from this approximation
is that radiative losses are much more important at the perpendicular
shock (where $\sigma\rs{B}=\sigma$) than at the parallel one (where
$\sigma\rs{B}=1$). In addition, the radiative losses depends rather
strongly on the index $\gamma$: $\sigma\rs{B}^2=16$ for $\gamma=5/3$
but $\sigma\rs{B}^2=49$ for $\gamma=4/3$.

{Note, that in the analysis above, the difference between the parallel 
and perpendicular shocks are only due to the compression factor $\sigma\rs{B}$ 
which may be treated as "compression-plus-amplification" factor. 
Therefore, our consideration may also be applied in case of the very 
turbulent and amplified pre-shock field (when information about obliquity is 
lost) once this factor is known. }

Our approximations reflects also the general `rule' for IC emission: there
is less IC emission where MF is stronger. Namely, the azimuthal variation
is $I\rs{ic}(\varphi)\propto 1-\mathrm{const}\times\psi(\varphi)/{\cal
F}(\varphi)^\alpha$ with $\psi\propto \sigma\rs{B}(\varphi)^2$: emitting
electrons disappears toward the shock with larger $\sigma\rs{B}$
because MF strength is a reason of higher losses there. Similar
dependence on $\sigma\rs{B}$ is for X-rays, Eq.~(\ref{xmaps:Ixray});
it is however dominated by the increased term $S\rs{x}(\varphi)\propto
\sigma\rs{B}(\varphi)^{3/2}$.

TeV \g-ray image of SN~1006 demonstrates good correlations with  X-ray
image smoothed to the H.E.S.S. resolution \citep{HESS-sn1006-2010}.
We mean here both the location and sizes of the bright limbs. Let us
consider the polar-caps model of SN~1006. The shock is quasi-parallel
around the limbs, MF azimuthally increases (in $\approx 4$ times).
{The injection efficiency decreases (in $>10^{-3}$ times) out of
the limbs; the number of electrons emitting in X-rays and TeV \g-rays
is therefore} dramatically low at perpendicular shock comparing to the parallel,
that is in agreement with no TeV emission at NW and SE regions of
SN~1006. However, the azimuthal sizes of the limbs in X-rays and \g-rays
{is expected} to be different, in the polar-caps model: they should be larger
in X-rays. Really, 
{azimuthal variation of ${\cal E}\rs{max}$ is smaller than variations of 
$\varsigma$ and $\sigma\rs{B}^{3/2}$; therefore,
from Eqs.~(\ref{xmaps:xray_brightness}) and (\ref{xmaps:IC_brightness}), 
azimuthal variations of brightness are mostly}
$S\rs{x}\propto \varsigma\sigma\rs{B}^{3/2}$
while $S\rs{ic}\propto \varsigma$ 
({see Fig.~\ref{xmaps:limb_sizes} for a comparison of remnant 
morphologies in the radio, X-ray and \g-ray bands}). 
We hope that future observations allow
us to see if there is a difference in azimuthal sizes of the limbs in
various bands.

How back-reaction of accelerated particles may modify nonthermal
images of SNRs? Our formulae can restore some of these effects.
In our approximations, $s$, in general, is allowed to vary with $E$,
e.g. to be $s(E)=s+\delta s(E)$. The index $s$ reflects the 'local'
slope of the electron spectrum around $\epsilon\rs{m}$. Therefore, if
$s(E)\neq\mathrm{const}$, the index $s$ has to vary with azimuth
because $\epsilon\rs{m}$ varies, Eq.~(\ref{xmaps:epmX}). Generally
speaking, such approach allows one to estimate the role of the nonlinear
`concavity' of the electron spectrum in modification of the nonthermal
images. However, we expect that this effect is almost negligible
because $\delta s(E)$ is very slow function, at least within interval
of electron energies contributing to images. Another effect of
efficient acceleration consists in the adiabatic index $\gamma$ smaller
than $5/3$. Our approximations are written for general $\gamma$. Namely,
the index $\gamma$ affects $S(\varphi,\bar\varrho)$ through $\sigma$.
Cosmic rays may also cause the amplification of the seed ISMF.  In our
formulae, $\sigma\rs{B}(\Theta\rs{o})$ represents the obliquity variation
of the ratio of the downstream MF to ISMF strength, $B\rs{s}/B\rs{o}$.
Therefore, it may account for both the compression and amplification of
ISMF; for such purpose, $\sigma\rs{B}$ should be expressed in a way to
be unity at parallel shock.

\section{Comparison with observations: a working example}

The main reason for the derivation of the explicit approximations
of Sect. \ref{xmaps:approximation_section} is to highlight the factors which are the
most efficient in the formation of the pattern of surface
brightness. However, sometimes the analytical formulae may help in
estimates some of the remnant parameters. We would like to present
two examples. Let us consider SN~1006, $s=2$ \citep{SN1006Marco},
$\gamma=5/3$.

\begin{figure}
 \centering
 \includegraphics[width=7.7truecm]{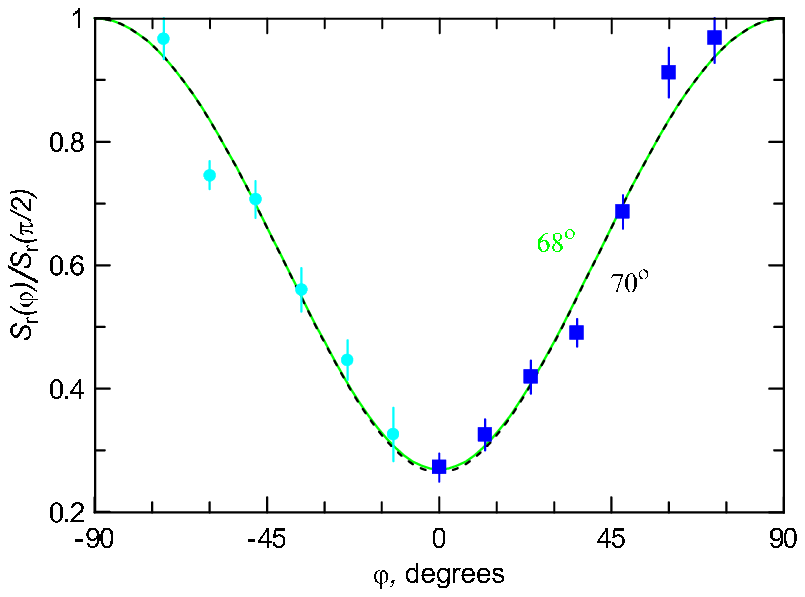}
 \caption{Azimuthal profiles of the radio brightness of SN~1006. Experimental data and the best-fit theoretical 
 profile calculated from detailed model (dashed line) are from Paper I. Green solid line represents the 
 best-fit profile obtained from the approximate analytical formula (\ref{xmaps:radio_brightness}). 
                }
 \label{xmaps:fig_radio_ap_sn1006}
\end{figure}

The injection efficiency is isotropic if one assumes that
SN~1006 evolves in the uniform ISMF and uniform ISM (Paper I).
The best-fit value of the aspect angle found from the approximate
Eq.~(\ref{xmaps:radio_brightness}) is $68^\mathrm{o}\pm 4.0^\mathrm{o}$
(Fig.~\ref{xmaps:fig_radio_ap_sn1006}) while the detailed numerical
calculations give $70^\mathrm{o}\pm 4.2^\mathrm{o}$ (Paper I).

The same approximate formula shows that the radial profile of radio
brightness depends only on the value of $b$ (which shows how the
injection efficiency evolve with the shock velocity).  Unfortunately,
the differences between profiles for $b=-1$ and $b=1$ are comparable
with accuracy of the approximate formulae and of the experimantal
data; therefore, the approximation may not be used for estimations of $b$.

\begin{figure}
 \centering
 \includegraphics[width=7.6truecm]{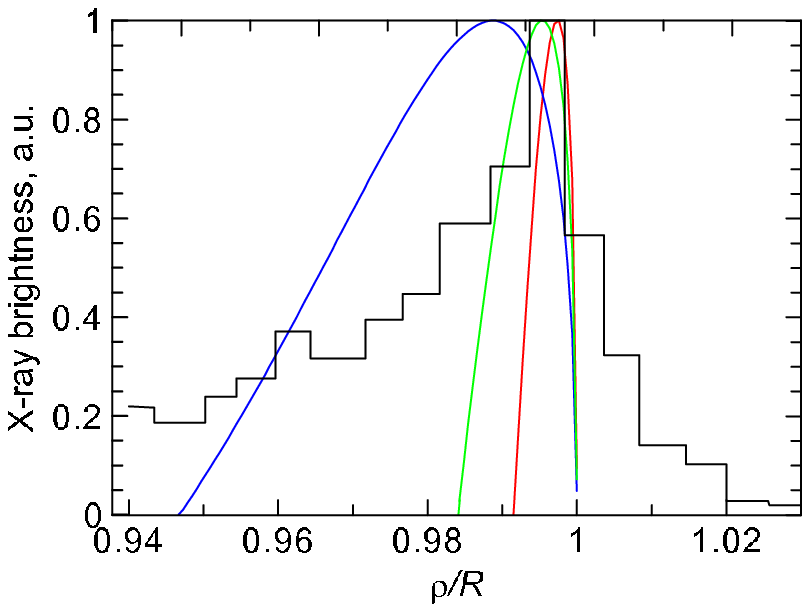}
 \caption{Radial profiles of the hard X-ray brightness of SN~1006. Experimental data 
 \citep[from Fig. 4A in][]{Long-et-al-2003} are shown by histogram. 
 Profiles given by approximation (\ref{xmaps:xray_brightness}) for 
 $B=50,100,150\un{\mu G}$ are shown by blue, green and red lines respectively. 
 Other parameters: $\varepsilon=1.2\un{keV}$ and $\nu\rs{break\|}=150\un{eV}$, 
 $\phi\rs{o}=68^\mathrm{o}$, $\varphi=70^\mathrm{o}$, 
 ${\cal E}\rs{max}=2.9$, $q=0$, 
 $b=0$, $\alpha=1$. 
                }
 \label{xmaps:fig_xray_ap_sn1006}
\end{figure}

The sharpest radial profile of X-ray brightness \citep[from Fig. 4A
in][]{Long-et-al-2003} was used to estimate the strength of the
post-shock MF in $\sim 100\un{\mu G}$ \citep{Ber-Volk-2003-mf}.
Fig.~\ref{xmaps:fig_xray_ap_sn1006} shows approximate profiles for three
values of MF in comparison with the Chandra profile and detailed numerical
simulations (Petruk et al., in preparation). One can also see from the
approximation that $B\sim 100\un{\mu G}$ is the most appropriate value
in agreement with the value found in the literature.

\section{Conclusions}
\label{xmaps:conclusion_section}

The present paper extends analysis of properties of the surface
brightness distribution of spherical adiabatic SNRs started in Paper
I (radio band) and Paper II (IC \g-rays) to the nonthermal X-rays. It
also generalizes the method of approximate analytical description of the
azimuthal and radial profiles of brightness introduced in these papers.

Synchrotron images of adiabatic SNR in X-rays are synthesized for
different assumptions about obliquity variations of the injection
efficiency, MF and maximum energy of accelerated electrons. We analyze
properties of these images. Different models of electron injection (quasi-parallel,
isotropic and quasi-perpendicular) as well as models of the electron
maximum energy (time-limited, loss-limited and escape-limited) are
considered.

The azimuthal variation of the synchrotron X-ray and IC \g-ray brightness
is mostly determined by variations of $\varsigma$, $\sigma\rs{B}$ and
$E\rs{max}$, of the radio brightness by $\varsigma$ and $\sigma\rs{B}$
only. In general, higher $B$ increases X-ray and decreases IC \g-ray
brightness.
Really, higher MF is a reason of larger losses of emitting electrons
(i.e. decrease of their number) and thus of the smaller brightness due
to IC process. In contrast, X-rays are more efficient there because
$S\rs{x}\propto B^{3/2}$.

The radial profiles of brightness depend on a number of factors. It is
quite sensitive to the adiabatic index: $\gamma<5/3$ makes plasma more
compressible. Therefore, the brightness profile is thinner due to larger
compression factor, larger gradient of density and MF downstream of the
shock and larger radiative losses.

The role and importance of various factors on the surface brightness
in radio, synchrotron X-rays and IC \g-rays are demonstrated by the
approximate analytical formulae. They accurately represent numerical
simulations close to the shock and are able to account for some non-linear
effects of acceleration if necessary.
This makes the approximations a powerful tool for quick analysis of the
surface brightness distribution due to emission of accelerated electrons
around SNR shells. The application of the approximate formulae
to the case of SN1006 yields measures of the aspect angle and the
post-shock MF in good 
agreement with more accurate analysis found
in the literature.

\appendix
\section[]{Evolution of the electron energy spectrum 
downstream of the adiabatic shock}
\label{xmaps:app2}

Relativistic electrons evolving downstream of the shock suffer from adiabatic expansion and radiative losses due to synchrotron and inverse Compton processes. 
Electrons are considered to be confined to the fluid element which removed them from the acceleration site \citep{Reyn-98}.%
\footnote{{Such an assumption means that we consider the advection only, not the diffusion. 
However, for X-ray emitting electrons, the lengthscales for advection $l\rs{adv}$ 
and diffusion $l\rs{dif}$ are comparable: $l\rs{adv}/l\rs{dif}=0.8-1.1$ \citep{Ballet-2006}. 
Therefore, the results of the present paper are robust.}}

Let the fluid element with Lagrangian coordinate $a$ was shocked at time $t\rs{i}$. 
If energy of electrons was $E\rs{i}$ at $t\rs{i}$, it becomes later \citep{Reyn-98} 
\begin{equation}
 E={E\rs{i} \over \bar{n}(\bar{a})^{-1/3}+{\cal I}(\bar{a}){E\rs{i}}/{E\rs{f}}}
 \label{energylosses}
\end{equation} 
where the first summand in the denominator reflects adiabatic losses, the second one is due to radiative losses, 
$\bar n=n/n\rs{s}$, index 's' means `immediately post-shock', the fiducial energy for parallel shock  $E\rs{f}=637/(B\rs{eff,s\|}^{2}t)\un{cgs}$. 
The effective magnetic field is ${B}\rs{eff}^2={B}^2+{B}\rs{CMB}^2$, {$B(a)$ takes into account its evolution downstream \citep{Reyn-98}}, $B\rs{CMB}=3.27\un{\mu G}$ strength of magnetic field with energy density equal to energy density of CMBR. $B\rs{CMB}$ is introduced in order to account for the inverse-Compton losses \citep{Reyn-98}, therefore it is constant everywhere. The synchrotron channel dominates inverse-Compton losses if $B\rs{s}\gg B\rs{CMB}$.

The dimensionless function ${\cal I}$ accounts for evolution of fluid during time from $t\rs{i}$ to $t$; 
it was initially defined as integral over time \citep{Reyn-98}. 
In case of Sedov shock, ${\cal I}$ may be written in terms of spatial coordinate  
that is more convenient for simulations than original representation in terms of time. Namely, 
for uniform ISM: 
\begin{equation}
 {\cal I}(\bar{a},\Theta\rs{o},d)=
 \frac{5\sigma\rs{B}^2}{2\bar{n}(\bar{a})^{1/3}}\int_{\bar{a}}^{1} 
 x^{3/2}{\bar{B}\rs{eff}\left({\bar{a}\over x}\right)^2}
 \bar{n}\left({\bar{a}\over x}\right)^{1/3}dx,
 \label{Int}
\end{equation} 

Eq.~(\ref{energylosses}) results in relations 
\begin{equation}
 E\rs{i}=\frac{E}{{\cal E}\rs{ad}{\cal E}\rs{rad}},
\qquad
 dE\rs{i}= \frac{dE}{{\cal E}\rs{ad}{\cal E}^2\rs{rad}}
\end{equation}
where the adiabatic and radiative losses are represented by
\begin{equation}
 {\cal E}\rs{ad}={\bar{n}(\bar{a})^{1/3},\qquad 
 {\cal E}\rs{rad}=1-{\cal I}(\bar{a},\Theta\rs{o})E/E\rs{f}}.
 \label{calE}
\end{equation}

Shocks of different strength are able to accelerate electrons to different $E\rs{max}$. 
Let $E\rs{max}\propto V^q$ where $V$ is the shock velocity, $q=1,2,0$ for loss-limited, time-limited and 
escape-limited models respectively \citep{Reyn-98}. 
If shock accelerates electrons to $E\rs{max}$ at present time $t$, then, at some previous time $t\rs{i}$ when 
fluid element $a\equiv R(t\rs{i})$ was shocked, the shock was able to accelerate electrons to 
\begin{equation}
 E\rs{max}(t\rs{i})=E\rs{max}\left(\frac{V(t\rs{i})}{V(t)}\right)^{q}
 =E\rs{max}\bar a^{-3q/2} ,
\end{equation}
where $\bar a=a/R(t)$ and we used Sedov solutions.
The obliquity variation of the maximum energy is given by 
$E\rs{max}=E\rs{max,\|}{\cal E}\rs{max}(\Theta\rs{o})$ with ${\cal E}\rs{max}(\Theta\rs{o})$ 
independent of time. 

Let us assume that, at time $t\rs{i}$, an electron distribution has been produced at the
shock  
\begin{equation}
 N(E\rs{i},t\rs{i})=K\rs{s}(t\rs{i},\Theta\rs{o})E\rs{i}^{-s}
 \exp\left[{-\left({E\rs{i}\over E\rs{max}(t\rs{i},\Theta\rs{o})}\right)^{\alpha}}\right] 
 \label{sp-ini}
\end{equation}
where $\alpha$ is constant. 
The obliquity variation of $K\rs{s}$ is given by 
$K\rs{s}(\Theta\rs{o})=K\rs{s\|}{\cal K}(\Theta\rs{o})$ with 
${\cal K}(\Theta\rs{o})$ also independent of time.

Conservation equation
\begin{equation}
 N(E,a,t)=N(E\rs{i},a,t\rs{i}){a^2dadE\rs{i}\over \sigma r^2drdE} 
 \label{cons-N}
\end{equation}
(where $\sigma=n\rs{s}/n\rs{o}$) 
and continuity equation
$n_{\rm o}(a)a^2da=n(a,t)r^2dr$ 
shows that downstream 
\begin{equation}
\begin{array}{l}
 N(E,a,t)=K(a,t,\Theta\rs{o})\ E^{-s}\  {\cal E}\rs{rad}(\bar{a},E,\Theta\rs{o})^{s-2}
 \\ \\ \displaystyle \ \ \ \ \times
  \exp\left[{-\left({E\ \bar a^{\ \!3q/2}\over E\rs{max\|}(t)\ {\cal E}\rs{ad}(\bar{a})\ {\cal E}\rs{rad}(\bar{a},E)\ 
{\cal E}\rs{max}(\Theta\rs{o})
  }\right)^{\alpha}}\right]
\end{array}  
\label{N-downstr}
\end{equation}
with $K(a,t)=K_{\rm s}(t\rs{i})\bar{n}{\cal E}\rs{ad}^{s-1}$. 
If $K_{\rm s}\propto V^{-b}$, then evolution of $K$ is self-similar downstream 
\begin{equation}
 \bar{K}(\bar a)=K(a,t)/K_{\rm s}(t)=
 \bar{a}^{\ \!3b/2}\ 
 \bar{n}(\bar{a})^{(2+s)/3}.
 \label{self-K}
\end{equation}
Therefore, in general, 
\begin{equation}
 K(a,t,\Theta\rs{o})=K\rs{s\|}(t){\cal K}(\Theta\rs{o})\bar{K}(\bar a)
\end{equation}
where the profile $\bar{K}(\bar a)$ is independent of obliquity. 
Note, that $K(a,t,\Theta\rs{o})$ is not affected by the radiative losses, therefore 
it behaves in the same way also for radio emitting electrons. 
Once $s$ is close to 2, the radiative losses influence the shape of $N(E)$ mostly 
through the exponential term in Eq.~(\ref{N-downstr}). In other words, 
they are effective only around the high-energy end of the electron spectrum as it is shown by \citet{Reyn-98}.

The above formulae are also valid if the spectral index $s$ depends on $E$, e.g. $s(E)=s+\delta s(E)$ like it would be in case of the nonlinear acceleration. 
In addition, no specific value of the adiabatic index $\gamma$ is assumed here. It influences the downstream evolution of relativistic electrons through $\bar n (\bar a)$ which depends on $\gamma$ \citep{Sedov-59}.

\section[]{Approximations for distributions of some parameters behind 
the adiabatic shock}
\label{xmaps:app3}

Let us find approximations for dependence of some parameter 
$\bar{\cal X}\equiv X/X(R)$ on the Lagrangian coordinate $\bar a\equiv a/R$ downstream close to the adiabatic shock. 
We are interested in approximations of the form
\begin{equation}
 \bar {\cal X}(\bar{a})\approx \bar{a}^{\ \kappa}
\end{equation}
where, by definition,
\begin{equation}
 \kappa= \left[-{a\over {\cal X}\rs{*}(a)}{\partial {\cal X}\rs{*}(a)\over \partial a}\right]_{a=R}
 			 = \left[-{\partial \ln {\cal X}\rs{*}(a)\over \partial \ln a}\right]_{a=R}
 \label{kappa-def}
\end{equation}
and star marks the dependence given by the Sedov solution. 

This approach yields for density
\begin{equation}
 \bar n(\bar a)\approx \bar a^{\kappa\rs{na}},\qquad \kappa\rs{na}=\frac{5\gamma+13}{(\gamma+1)^2},
\end{equation} 
for the relation between Eulerian and Lagrangian coordinates
\begin{equation}
 \bar r\approx \bar a^{1/\sigma},\qquad \bar r\rs{\bar a}\approx (1/\sigma)\bar a^{(1/\sigma)-1}
\end{equation} 
where the shock compression factor is
\begin{equation}
 \sigma=\frac{\gamma+1}{\gamma-1}.
\end{equation} 
Note that the density distribution in Eulerian coordinates is much more sensitive to $\gamma$ (Table~\ref{xmaps:table_param_gamma}):
\begin{equation}
  \bar n(\bar r)\approx \bar r^{\kappa\rs{nr}},\qquad \kappa\rs{nr}=\sigma\kappa\rs{na}=\frac{5\gamma+13}{(\gamma+1)(\gamma-1)}.
\end{equation}

Magnetic field is approximately
\begin{equation}
 \bar B\approx \bar a^{\beta(\Theta\rs{o,eff})}, 
\end{equation} 
\begin{equation}
 \beta(\Theta\rs{o,eff})=
      \frac{\beta_\|\cos^2\Theta\rs{o}+\beta_\bot\sigma^2\sin^2\Theta\rs{o}}
              {\cos^2\Theta\rs{o}+\sigma^2\sin^2\Theta\rs{o}},
 \label{xmaps:appE-beta}
\end{equation} 
\begin{equation}
 \beta_\|=\frac{4}{\gamma+1}, 
 \qquad
 \beta_\bot=\frac{3\gamma+11}{(\gamma+1)^2}.
\end{equation} 
Approximation for normalization $K$ follows from the definition (Appendix \ref{xmaps:app2})
\begin{equation}
 \bar K=\bar a^{3b/2}\bar n^{(2+s)/3}
\end{equation} 
and approximation for $\bar n$.

Adiabatic losses are accounted with ${\cal E}\rs{ad}(a)$ which is defined by (\ref{calE}). 
Its approximation is therefore 
\begin{equation}
 {\cal E}\rs{ad}(\bar a)\approx\bar{a}^{\kappa\rs{ad}},
 \qquad 
 \kappa\rs{ad}=\frac{5\gamma+13}{3(\gamma+1)^2},
 \label{calEad-approx}
\end{equation} 
it is valid for $\bar{r}>0.8$ with error less than few per cent. The value of $\kappa\rs{ad}$ is close to unity for $\gamma=1.1\div 5/3$ (Table~\ref{xmaps:table_param_gamma}). 

In order to approximate ${\cal E}\rs{rad}$ defined by (\ref{calE}), we substitute (\ref{Int}) with approximations for $\bar n$ and $\bar B$. 
Then we use the property 
\begin{equation}
 \lim\limits_{a\rightarrow 1}\left(\frac{a}{f(a)}\frac{df}{da}\right)=cy
\end{equation}
for function of the form $f(a)=1-ca^x(1-a^y)$. In this way,  
\begin{equation}
 {\cal E}\rs{rad}(\bar a,E)\approx \bar{a}^{\kappa\rs{rad}}, \qquad \kappa\rs{rad}=\frac{5\sigma\rs{B}^2(\Theta\rs{o})E}{2E\rs{f\|}}.
 \label{calErad-approx}
\end{equation}
This expression is good for $\bar{r}>0.94$, with error of few per cent. It depends on $\gamma$ through $\sigma$ in $\sigma\rs{B}$ which is \citep{Reyn-98}
\begin{equation}
 \sigma\rs{B}=\left(\frac{1+\sigma^2 \tan^2\Theta\rs{o}}{1+\tan^2\Theta\rs{o}}\right)^{1/2}.
\end{equation} 

Note that dependence on the absolute value of the magnetic field strength $B\rs{s\|}$ is present in (\ref{calErad-approx}): $E\rs{f\|}\propto B\rs{s\|}^{-2}$. This approximation clearly shows that the fiducial energy $E\rs{f}$ is important parameter reflecting the `sensitivity' to the radiative losses, as it shown by \citet{Reyn-98}: the larger the fiducial energy the smaller the radiative losses. 
In fact, ${\cal E}\rs{rad}=1$ means no radiative losses at all, see (\ref{calE}). Another fact directly visible from Eq.~(\ref{calErad-approx}) is that radiative losses are much more important at the perpendicular shock ($\sigma\rs{B}=\sigma$) than at the parallel one ($\sigma\rs{B}=1$). Radiative losses depends rather strongly on the index $\gamma$: $\sigma\rs{B}^2=16$ for $\gamma=5/3$ but $\sigma\rs{B}^2=49$ for $\gamma=4/3$. 

The values of parameters in approximations for different adiabatic index $\gamma$ are presented in Table~\ref{xmaps:table_param_gamma}.

\begin{table}	
 \caption{Parameters in approximations}
 \begin{tabular}{lccc}
	\hline
	       Expression & $\gamma=5/3$ & $\gamma=4/3$ & $\gamma=1.1$ \\
	\hline
	 $\kappa\rs{na}=\displaystyle\frac{5\gamma+13}{(\gamma+1)^2}$ & 3 & 3.6 & 4.2\\[8pt]
	 $\kappa\rs{nr}=\displaystyle\frac{5\gamma+13}{(\gamma+1)(\gamma-1)}$ & 12 & 25& 88\\[8pt]
	 $\sigma=\displaystyle\frac{\gamma+1}{\gamma-1}$ & 4 & 7 & 21 \\[8pt]
   $\beta\rs{\|}=\displaystyle\frac{4}{\gamma+1}$ &1.5&1.7&1.9\\[8pt]
	 $\beta\rs{\bot}=\displaystyle\frac{3\gamma+11}{(\gamma+1)^2}$ &2.2&2.8&3.2\\[8pt]
	$\kappa\rs{ad}={\kappa\rs{na}}/{3}$&1&1.2&1.4\\
	\hline
 \end{tabular}
 \label{xmaps:table_param_gamma}
\end{table}

\section[]{Approximate formula for the azimuthal variation of the synchrotron X-ray surface brightness in Sedov SNR}
\label{xmaps:app1}

A formula obtained here may be useful in situations when an approximate quantitative estimation for the azimuthal variation of the synchrotron X-ray surface brightness is sufficient. 

{\bf 1.} 
The emissivity due to synchrotron emission is
\begin{equation}
 q(\varepsilon)=\int  N(E) p(E,\varepsilon)dE
 \label{xmaps:eq5} 
\end{equation}
Spectral distribution of the synchrotron radiation power of electrons with energy 
$E$ in magnetic field of the strength $B$ is 
\begin{equation}
 p(E,\nu)={\sqrt{3}e^3B\sin\phi\over m\rs{e}c^2}F\left({\nu\over\nu\rs{c}}\right),  
 \label{xmaps:eq6}
\end{equation}
where $\nu$ is frequency, $\nu\rs{c}=c_1BE^2$ the characteristic frequency.
Most of this radiation is in photons with energy 
$\varepsilon\rs{p}=0.29h\nu\rs{c}$. 
In the 'delta-function approximation', the special function $F$ is substituted with
\begin{equation} 
 F\left(\frac{\nu}{\nu\rs{c}}\right)=
 \delta\left(\frac{\nu}{\nu\rs{c}}-0.29\right)
 \int\limits_{0}^{\infty}F(x)dx
 \label{F-delta}
\end{equation}
where
\begin{equation} 
 \int\limits_{0}^{\infty}F(x)dx=\frac{8\pi}{9\sqrt{3}}.
\end{equation} 
With this approximation, (\ref{xmaps:eq5}) becomes 
\begin{equation}
 q(\varepsilon)=\frac{4\pi e^3\sin\phi\ \varepsilon^{1/2}B^{1/2}}{9m\rs{e}c^2\ 0.29c_1^{1/2}h^{1/2}}N(E\rs{m})
 \label{xmaps:q} 
\end{equation}
where $E\rs{m}$ is the energy of electrons which give maximum contribution to synchrotron emission 
at photons with energy $\varepsilon$: 
$E\rs{m}={\varepsilon}^{1/2}\left({0.29hc_1B}\right)^{-1/2}$.

{\bf 2.} 
Let the energy of relativistic electrons is $E$ in a given fluid element at present time. 
Their energy was $E\rs{i}$ at the time this element was shocked. 
These two energies are related as 
\begin{equation}
 E=E\rs{i}{\cal E}\rs{ad}{\cal E}\rs{rad}
\end{equation}
where ${\cal E}\rs{ad}$ accounts for the adiabatic losses and ${\cal E}\rs{rad}$ for the radiative losses
(Appendix \ref{xmaps:app2}). 
There are approximations valid close to the shock (Appendix \ref{xmaps:app3}): 
\begin{equation} 
 {\cal E}\rs{ad}\approx \bar a^{\kappa\rs{ad}},\qquad {\cal E}\rs{rad}\approx \bar a^{5\sigma\rs{B}^2E/2E\rs{f,\|}}
 \label{approxE}
\end{equation}
where $\bar a=a/R$, $a$ is Lagrangian coordinate of the fluid element, 
$E\rs{f,\|}$ is the fiducial energy for parallel shock, $\kappa\rs{ad}$ depends on $\gamma$ and is given by (\ref{calEad-approx}); $\kappa\rs{ad}=1$ for $\gamma=5/3$ (for other $\gamma$ see Table~\ref{xmaps:table_param_gamma}).
The factor 
$\sigma\rs{B}$ represents compression in the classical MHD \citep{Reyn-98} 
but may be interpreted also as amplification-plus-compression factor. 
In the latter case, it should be written in a way to be unity at parallel shock.

The downstream evolution of $K$ in a Sedov SNR is (Appendix \ref{xmaps:app2})
\begin{equation}
 K\propto\varsigma(\Theta\rs{o})\bar K(\bar a)
\end{equation}
where $\varsigma$ is injection efficiency. 
With the approximations (\ref{approxE}) and $s$ close to 2, the distribution  
$N(E)$ may be written from (\ref{N-downstr}) as
\begin{equation}
 N(E,\Theta\rs{o})\propto \varsigma(\Theta\rs{o})\bar K(\bar a)E^{-s}
 \exp\left[-\left(\frac{E\bar a^{-\psi(E,\Theta\rs{o})}}{E\rs{max,\|}{\cal E}\rs{max}(\Theta\rs{o})}\right)^{\alpha}\right]
 \label{xmaps:N_E_s}
\end{equation}
where 
\begin{equation}
 \psi(E,\Theta\rs{o})=\kappa\rs{ad}+\frac{5\sigma\rs{B}(\Theta\rs{o})^2E}{2E\rs{f,\|}}-\frac{3q}{2}
\end{equation}
and $s$ is allowed to vary with $E$. 

{\bf 3.} 
Let us consider the azimuthal profile of the synchrotron X-ray brightness $S\rs{\varrho}$ 
at a given radius $\varrho$ from the center of the SNR projection. 

Like in Paper II, we consider the `effective' obliquity angle $\Theta\rs{o,eff}$ which,
for a given azimuth, equals to the obliquity angle for a sector with the same azimuth 
in the plane of the sky (see details in Paper II). 
The relation between the azimuthal
angle $\varphi$, the obliquity angle $\Theta\rs{o,eff}$ and the aspect angle
$\phi\rs{o}$ is as simple as
\begin{equation}
 \cos\Theta\rs{o,eff}\left(\varphi,\phi\rs{o}\right)=\cos\varphi\sin\phi\rs{o}
\end{equation}
for the azimuth angle $\varphi$ measured from the direction of ISMF in the
plane of the sky.

The surface brightness of SNR projection at distance $\varrho$ from the center and at azimuth $\varphi$ is 
\begin{equation}
 S(\bar\varrho,\varphi)=2\int^{1}_{\bar a(\bar\varrho)}q(\bar a) {\bar r \bar r\rs{\bar a} d\bar a\over 
 \sqrt{\bar r^2-\bar \varrho^2}}.
\end{equation}
where $\bar r\rs{\bar a}$ is the derivative of $\bar r(\bar a)$ in respect to $\bar a$.
The azimuthal variation of the synchrotron X-ray brightness 
is approximately 
{\small
\begin{equation}
\begin{array}{ll}
 S\rs{x}&\propto
 \displaystyle
 \varsigma(\Theta\rs{o,eff})\sigma\rs{B}(\Theta\rs{o,eff})^{(s+1)/2}
 I\rs{rx}(\Theta\rs{o,eff},\bar\varrho)
 \\ \\ &\times\displaystyle
 \exp\left[-\left(\frac{E\rs{ms}(\varepsilon,\Theta\rs{o,eff})}
 {E\rs{max,\|}{\cal E}\rs{max}(\Theta\rs{o,eff})}\right)^{\alpha}\right]
\end{array}
 \label{Xazimuth:eq2}
\end{equation}
where 
\begin{equation}
\begin{array}{ll}
 I\rs{rx}&=
 \displaystyle
 \int^{1}_{\bar a(\bar\varrho)}{\bar K \bar B^{(s+1)/2} \bar r \bar r\rs{\bar a} \over 
 \sqrt{\bar r^2-\bar \varrho^2}} 
 \\ \\ &\times\displaystyle
 \exp\left[-\left(\frac{E\rs{ms}}
 {E\rs{max,\|}{\cal E}\rs{max}}\right)^{\alpha}\left(\bar a^{-\alpha\psi(E\rs{m})}\bar B^{-\alpha/2}-1\right)\right]
 d\bar a,
\end{array}
 \label{Xazimuth:eq2:int}
\end{equation}
} \noindent 
reflects the dependence on $\rho$, 
$E\rs{ms}$ is $E\rs{m}$ for $\bar B=1$:
\begin{equation}
 E\rs{ms}(\varepsilon,\Theta\rs{o,eff})=
 \left(\frac{\varepsilon}{0.29hc_1B\rs{o}\sigma\rs{B}(\Theta\rs{o,eff})}\right)^{1/2}.
 \label{xmaps:Em}
\end{equation}
Note, that $E\rs{ms}\propto \varepsilon^{1/2}$, 
i.e. $S\rs{\varrho}$ depends in our approximation 
on the energy $\varepsilon$ of observed X-ray photons. 

{\bf 4.} 
Let us approximate $I\rs{rx}$. 
First, we use the approximations 
$\bar a\approx \bar r^{\sigma}$, $\bar K\bar B^{(s+1)/2}\bar r\rs{\bar a}\approx \bar a^{\kappa\rs{r}}/\sigma$, 
which are valid close to the shock (Appendix \ref{xmaps:app3}), $\sigma$ is the shock compression ratio. 
Next, we expand $\bar r/\sqrt{\bar r^2-\bar \varrho^2}$ in powers of the small parameter $(r-1)$ and consider the only first term of the decomposition:
\begin{equation}
 {\bar r\over\sqrt{\bar r^2-\bar \varrho^2}}\approx {1\over \sqrt{1-\bar \varrho^2}}.
\end{equation}
The exponential term in the integral expands in powers of the small parameter 
$(1-a)$: 
\begin{equation}
 \exp\left(-x_1(a^{-x_2}-1)\right)\approx 1-x_1x_2(1-a).
 \label{xmaps:exp_decomp1}
\end{equation}
In addition, $E\rs{ms}$ is used instead of $E\rs{m}$.

Close to the shock, the integral of interest is therefore 
\begin{equation}
 I\rs{rx}(\varphi,\bar\varrho)\approx 
 I\rs{r}(\bar \rho)I\rs{x}(\varphi,\bar\varrho)
\label{xmaps:Ix}
\end{equation}
where 
\begin{equation}
 I\rs{r}=
 {1\over \sigma\sqrt{1-\bar \varrho^2}} 
 {1-\bar\varrho^{\sigma(\kappa\rs{r}+1)}\over \kappa\rs{r}+1},
 \label{xmaps:app-d:i}
\end{equation}
\begin{equation}
 I\rs{x}=
 \left[1-\frac{\epsilon\rs{m}^{\alpha}(\psi+\beta/2)\alpha}{{\cal E}\rs{max}^{\alpha}}
 \left(1-\frac{1-\bar\varrho^{\sigma(\kappa\rs{r}+2)}}{1-\bar\varrho^{\sigma(\kappa\rs{r}+1)}}
 \frac{\kappa\rs{r}+1}{\kappa\rs{r}+2}\right)
 \right].
\end{equation}
The parameter
\begin{equation}
 \psi=\kappa\rs{ad}+\frac{5\sigma\rs{B}^{2}\epsilon\rs{m}}{2\epsilon\rs{f\|}}-\frac{3q}{2}
\end{equation}
is responsible for the losses of emitting electrons and the time evolution of $E\rs{max}$ on the shock. 
The value of $\kappa\rs{ad}$ is rather close to unity for $\gamma=1.1\div 5/3$ (Table~\ref{xmaps:table_param_gamma}); 
unless radiative losses (the second term in $\psi$) are negligible, one may use $\kappa\rs{ad}\approx 1$ for any $\gamma$.
Other parameters are
\begin{equation}
 \epsilon\rs{m}=\frac{E\rs{ms}}{E\rs{max,\|}}=\left(\frac{\tilde\varepsilon}{0.29\sigma\rs{B}}\right)^{1/2},
 \label{xmaps:ep_m_X}
\end{equation}
$\beta$ is given by Eq.~(\ref{xmaps:appE-beta}), 
\begin{equation}
 \kappa\rs{r}=\frac{3b}{2}+\frac{2+s}{3}\kappa\rs{na}+\frac{s+1}{2}\beta+\frac{1}{\sigma}-1.
\end{equation}

Parameters $\psi$, $\epsilon\rs{m}$, $\sigma\rs{B}$, ${\cal E}\rs{max}$ and $\beta$ depend on $\Theta\rs{o,eff}$ and therefore on the aspect angle $\phi\rs{o}$ and the azimuth angle $\varphi$. 

The parameter $\beta$ reflects differences between MF distribution downstream the shock of the different obliquity. It varies from $\beta\rs{\|}$ at parallel shock to $\beta\rs{\bot}$ at perpendicular one, Eq.~(\ref{xmaps:appE-beta}). In the approximate formulae, it appears in the combination $\beta/2$; the role of $\beta\in[\beta\rs{\|};\beta\rs{\bot}]$ is minor in modification of the approximate azimuthal and radial profiles. Therefore, in order to simplify the approximation, we may take $\beta/2\approx 1$. 


The index $s$ in (\ref{xmaps:N_E_s}), in general, is allowed to vary with $E$, e.g. to be $s(E)=s+\delta s(E)$. In our approximation, due to (\ref{F-delta}), $s$ reflects the 'local' slope of the electron spectrum appropriate to $\epsilon\rs{m}$. Therefore, if one assumes $s(E)\neq\mathrm{const}$, the index $s(\epsilon\rs{m})$ may vary with azimuth because $\epsilon\rs{m}$ varies, Eq.~(\ref{xmaps:ep_m_X}). 

{\bf 5.} 
The final formula is
\begin{equation}
 S\rs{x}(\varphi,\bar\varrho)\propto
 \varsigma(\varphi)\sigma\rs{B}(\varphi)^{(s+1)/2}
 \exp\left[-\left(\frac{\epsilon\rs{m}(\varphi)}
 {{\cal E}\rs{max}(\varphi)}\right)^{\alpha}\right]
 I\rs{rx}(\varphi,\bar\varrho;\epsilon\rs{f\|})
 \label{Xazimuth:eq2_fin}
\end{equation}
where only $I\rs{x}$ depends on $\bar\varrho$ and $\epsilon\rs{f\|}$. 

\begin{figure*}
\centering
\includegraphics[angle=270,width=17.6truecm]{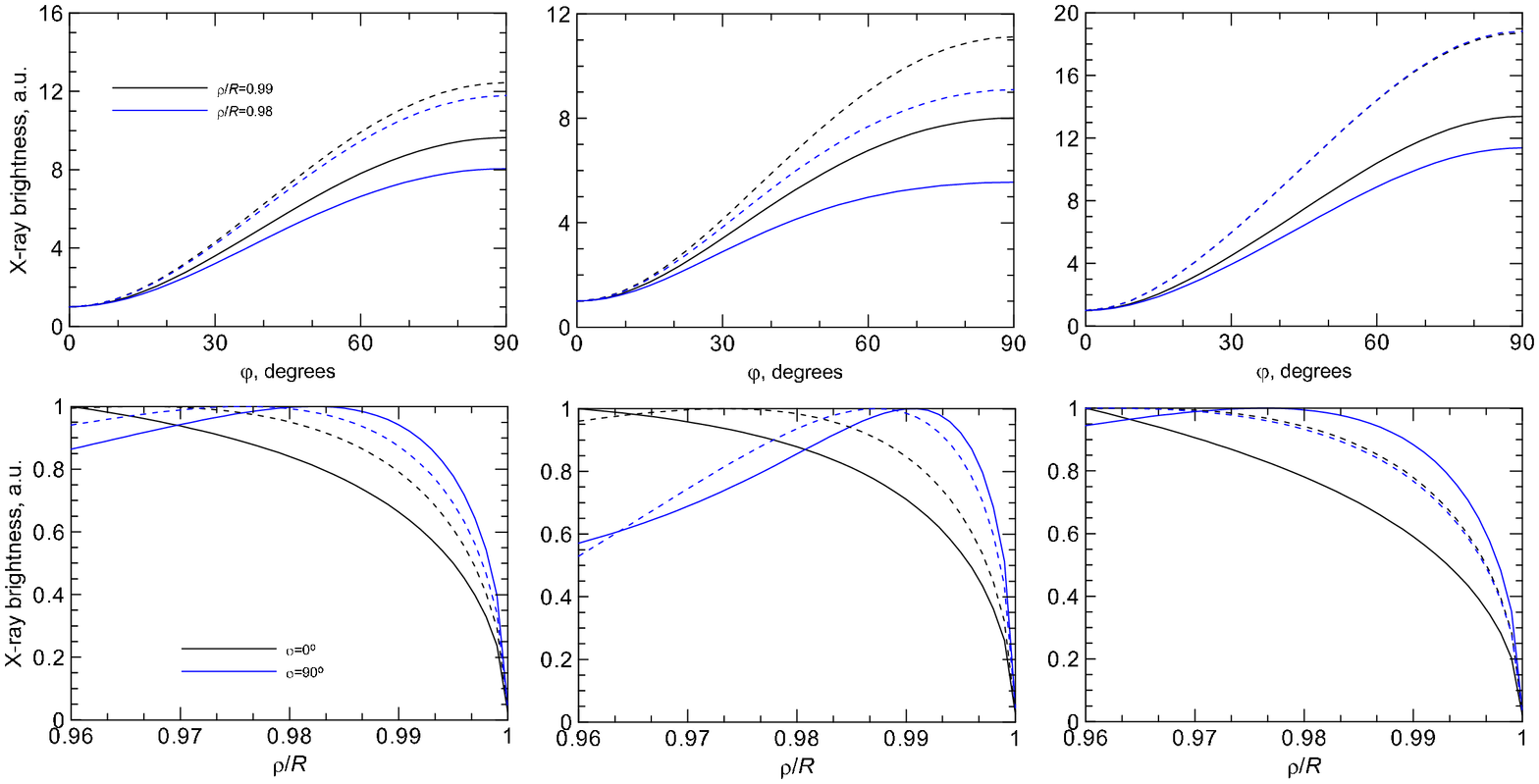} 
\caption{Azimuthal (upper panels) and radial (lower panels) profiles of the X-ray surface brightness $S\rs{x}$ (solid lines) and its approximations 
(\ref{Xazimuth:eq2_fin}) (dahsed lines). Calculations are done for $\phi\rs{o}=90^\mathrm{o}$, $b=0$, isotropic injection, $\gamma=5/3$, $s=2$, $\alpha=1$. 
Models of $E\rs{max}$: ${\cal E}\rs{max}=\mathrm{const}$ (left and middle panels) and time-limited one with $\eta=1.5$ (right panels). 
The reduced electron energy is $\epsilon\rs{m}=1$ and the reduced fiducial energy is $\epsilon\rs{f\|}=3$ (left), $\epsilon\rs{f\|}=1$ (middle), 
$\epsilon\rs{f\|}=5$ (right panels).
              } 
\label{xmaps:app-d-X}
\end{figure*}

\begin{figure*}
\centering
\includegraphics[angle=270,width=17.6truecm]{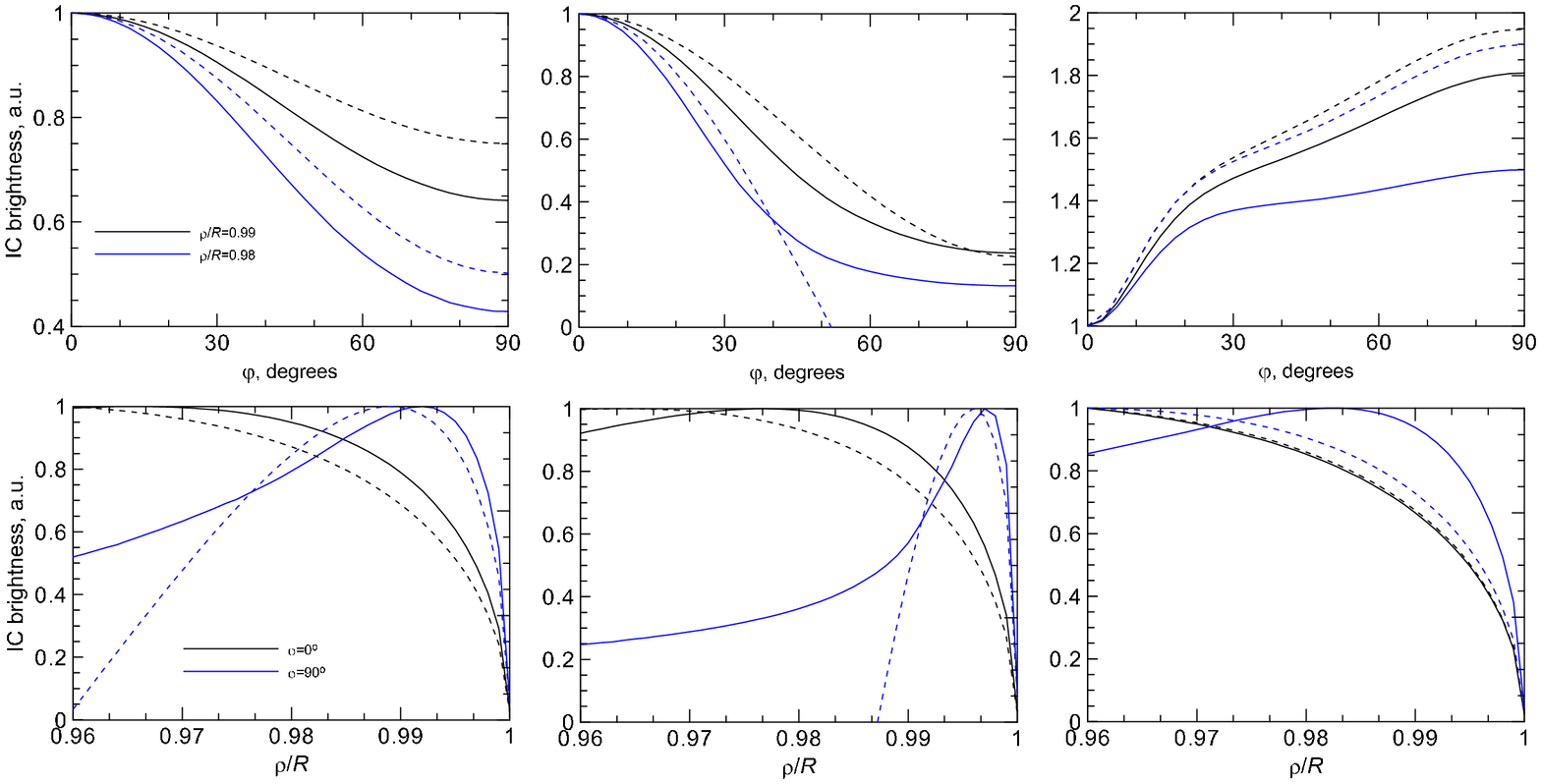} 
\caption{Azimuthal (upper panels) and radial (lower panels) profiles of the IC surface brightness $S\rs{ic}$ (solid lines) and its approximations (\ref{xmaps:ICazimuth}) (dahsed lines). Calculations are done for $\phi\rs{o}=90^\mathrm{o}$, $b=0$, isotropic injection, $\gamma=5/3$, $s=2$, $\alpha=1$. 
Models of $E\rs{max}$: ${\cal E}\rs{max}=\mathrm{const}$ (left and middle panels) and time-limited one with $\eta=1.5$ (right panels). 
The reduced electron energy is $\epsilon\rs{m}=1$ and the reduced fiducial energy is $\epsilon\rs{f\|}=3$ (left), $\epsilon\rs{f\|}=1$ (middle), $\epsilon\rs{f\|}=5$ (right panels).
              } 
\label{xmaps:app-d-IC}
\end{figure*}

The formula Eq.~(\ref{Xazimuth:eq2_fin}) gives us the possibility to approximate both the azimuthal and the radial brightness profiles of X-ray brightness for $\bar \varrho$ close to unity. 
It may be used 
(with a bit larger errors compared to the case of IC emission; Fig.~\ref{xmaps:app-d-X}, cf. Fig.~\ref{xmaps:app-d-IC}), 
for those azimuth $\varphi$ where $\epsilon\rs{m}\lsim 1$ and $\epsilon\rs{f}\gsim 0.1$, in the range of $\bar\varrho$ from $1-2\Delta\bar\varrho\rs{m}$ to 1, where $\Delta\bar\varrho\rs{m}=1-\bar\varrho\rs{m}$, $\bar\varrho\rs{m}$ is the radius where the maximum in the radial profile of brightness happens. We have in mind the maximum which is close to the shock, say $\bar\varrho\rs{m}>0.95$; therefore, in order to determine $\bar\varrho\rs{m}$, one should look for the azimuth with the largest radiative losses. This is discussed in details on example of the IC emission in Sect.~\ref{xmaps:app4}. 

Adiabatic index $\gamma$ affects the approximation through $\sigma$, $\kappa\rs{r}$, $\kappa\rs{ad}$. 

\section[]{Approximate formula for 
the IC gamma-ray surface brightness in Sedov SNR}
\label{xmaps:app4}

In Paper II, we have developed an analytic approximation for the azimuthal variation of the surface brightness of Sedov SNR in \g-rays due to the inverse-Compton process, for regions close to the forward shock. The approximation, Eq.~(11), in the cited paper accounts to zeroth order. However, like in the case of the X-ray brightness, the fall of the \g-ray emissivity downstream of the shock is quite strong in case of the efficient radiative losses of electrons. Therefore, in such cases of the efficient losses we need to consider the next order of approximation. 

Adopting the approach from the Appendix \ref{xmaps:app1} to IC emission (see also some details in Paper II), we come to the approximation 
\begin{equation}
 S\rs{ic}(\varphi)\propto
 \varsigma(\Theta\rs{o,eff})
 \exp\left[-\left(\frac{E\rs{m}(\varepsilon)}
 {E\rs{max,\|}{\cal E}\rs{max}(\Theta\rs{o,eff})}\right)^{\alpha}\right]
 I\rs{ic}(\Theta\rs{o,eff}, \bar\rho)
 \label{xmaps:ICazimuth0}
\end{equation}
where the energy $E\rs{m}$ of electrons which gives maximum contribution to IC emission at photons with energy $\varepsilon$ is 
\citep[e.g.][]{Pet08IC}
\begin{equation}
 E\rs{m}=\frac{m\rs{e}c^2\varepsilon^{1/2}}{2(kT)^{1/2}},
\end{equation} 
$T$ is the temperature of the black-body photons. 

The factor 
\begin{equation}
 I\rs{ic}=
  \int^{1}_{\bar a(\bar\varrho)}{\bar K \bar r \bar r\rs{\bar a} d\bar a\over 
 \sqrt{\bar r^2-\bar \varrho^2}} 
 \exp\left[-\left(\frac{E\rs{m}}
 {E\rs{max,\|}{\cal E}\rs{max}}\right)^{\alpha}\left(\bar a^{-\alpha\psi}-1\right)\right]
\end{equation} 
is approximately
\begin{equation}
\begin{array}{l}
I\rs{ic}(\varphi,\bar\varrho)\approx
 \displaystyle
 {1\over \sigma\sqrt{1-\bar \varrho^2}} 
 {1-\bar\varrho^{\sigma(\kappa\rs{ic}+1)}\over \kappa\rs{ic}+1}
 \\ \\ \times\displaystyle
 \left[1-\frac{\epsilon\rs{m}^{\alpha}\psi\alpha}{{\cal E}\rs{max}^{\alpha}}
 \left(1-\frac{1-\bar\varrho^{\sigma(\kappa\rs{ic}+2)}}{1-\bar\varrho^{\sigma(\kappa\rs{ic}+1)}}
 \frac{\kappa\rs{ic}+1}{\kappa\rs{ic}+2}\right)
 \right].
\end{array}
\label{xmaps:Ig}
\end{equation}
where $\kappa\rs{ic}$ and $\sigma$ comes from the approximations $\bar a\approx \bar r^{\sigma}$, $\bar K\bar r\rs{\bar a}\approx \bar a^{\kappa\rs{ic}}/\sigma$,
\begin{equation} 
 \psi=\kappa\rs{ad}+\frac{5\sigma\rs{B}^{2}\epsilon\rs{m}}{2\epsilon\rs{f\|}}-\frac{3q}{2},
\end{equation}
\begin{equation} 
 \epsilon\rs{m}=\frac{E\rs{m}}{E\rs{max,\|}}
 =\frac{\varepsilon^{1/2}}{2(kT)^{1/2}\gamma\rs{max\|}},
\end{equation}
\begin{equation} 
 \kappa\rs{ic}=\frac{3b}{2}+\frac{2+s}{3}\kappa\rs{na}+\frac{1}{\sigma}-1.
\end{equation}

The final formula is
\begin{equation}
 S\rs{ic}(\varphi,\bar\varrho)\propto
 \varsigma(\varphi)
 \exp\left[-\left(\frac{\epsilon\rs{m}}{{\cal E}\rs{max}(\varphi)}\right)^{\alpha}\right]
 I\rs{ic}(\varphi,\bar\varrho;\epsilon\rs{f\|})
 \label{xmaps:ICazimuth}
\end{equation}
It gives us the possibility to approximate both the azimuthal and the radial brightness profiles for $\bar \varrho$ close to unity. 

\subsection[]{Accuracy of the approximation}
\label{xmaps:app4_accuracy}

Fig.~\ref{xmaps:app-d-IC} demonstrates accuracy of the approximation (\ref{xmaps:ICazimuth}) (left and middle panels show in fact the variation of $I\rs{ic}$ because both $\varsigma$ and ${\cal E}\rs{max}$ are constant there). Our calculations may be summarized as follows: this approximation may be used, with errors less than $\sim 30\%$, for those azimuth $\varphi$ where $\epsilon\rs{m}\lsim 1$ and $\epsilon\rs{f}\gsim 0.1$, in the range of $\bar\varrho$ from $1-2\Delta\bar\varrho\rs{m}$ to 1, where $\Delta\bar\varrho\rs{m}=1-\bar\varrho\rs{m}$, $\bar\varrho\rs{m}$ is the radius (close to the shock) where the maximum in the radial profile of brightness happens; in addition, approximation may not be used for $\bar\varrho\lsim0.9$. If for some azimuth, the above 
conditions on $\epsilon\rs{m}$ and $\epsilon\rs{f}$ do not hold, the accuracy of approximation gradually decreases because the role of the exponent in $N(E)$ and of the radiative losses may not be described by the first terms in the decompositions used for derivation of the formula. 

Let's consider Fig.~\ref{xmaps:app-d-IC}. 
The photon energy $\epsilon\rs{m}$ does not change with azimuth for IC process. 
On the left panels, the reduced fiducial energy $\epsilon\rs{f}(\varphi)=\epsilon\rs{f\|}/({\cal E}\rs{max}\sigma\rs{B}^2)\gsim 0.1$ for any azimuth:
$\epsilon\rs{f\|}=3$ at the parallel shock and $\epsilon\rs{f\bot}=0.19$ at the perpendicular shock. 
The approximation is accurate for any azimuth, for $0.98\lsim\bar\varrho\leq 1$ at $\varphi=90^\mathrm{o}$ and for a wider range of $\varrho$ at $\varphi=0^\mathrm{o}$. Middle panels on Fig.~\ref{xmaps:app-d-IC} show the same case except of $\epsilon\rs{f\|}=1$. At parallel shock (i.e. $\varphi=90^\mathrm{o}$), the range for $\bar\varrho$ is smaller, $0.99\lsim\bar\varrho\leq 1$ (lower panel). Therefore, the approximation of the azimuthal profile for $\bar\varrho=0.98$ is inaccurate (upper panel, blue line), especially for $\varphi>45^\mathrm{o}$ where $\epsilon\rs{f}$ decreases; it is $\epsilon\rs{f\bot}=0.06$. The azimuthal profile is however accurate for $\bar\varrho=0.99$ (black line). Similar situation is for variable $E\rs{max}$ (right panels on Fig.~\ref{xmaps:app-d-IC}). $E\rs{max\bot}/E\rs{max\|}=3.25$ for considered model, therefore $\epsilon\rs{f\|}/\epsilon\rs{f\bot}=52$. Therefore, in order to obtain a representative approximation, the lowest possible $\epsilon\rs{f\|}$ should be about $0.1\times 52=5.2$. We see from the figure that accuracy decreases toward smaller $\epsilon\rs{f}$ (i.e. where the role of radiative losses are very efficient in modification of the electron distribution) and for smaller $\bar\varrho$. 

In general, the accuracy of the approximation is better for larger $\epsilon\rs{f}$ and smaller $\epsilon\rs{m}$. With decreasing of the aspect angle $\phi\rs{o}$, the accuracy of the approximations for the azimuthal profile increases at the beginning (because contrasts in $\sigma\rs{B}$, $\epsilon\rs{f}$ and ${\cal E}\rs{max}$ are lower) and then decreases again, for the case of the quasi-parallel injection, because SNR becomes centrally-brightened while our approximation is developed for regions close to the edge of SNR.

\begin{figure}
\centering
\includegraphics[width=6.2truecm]{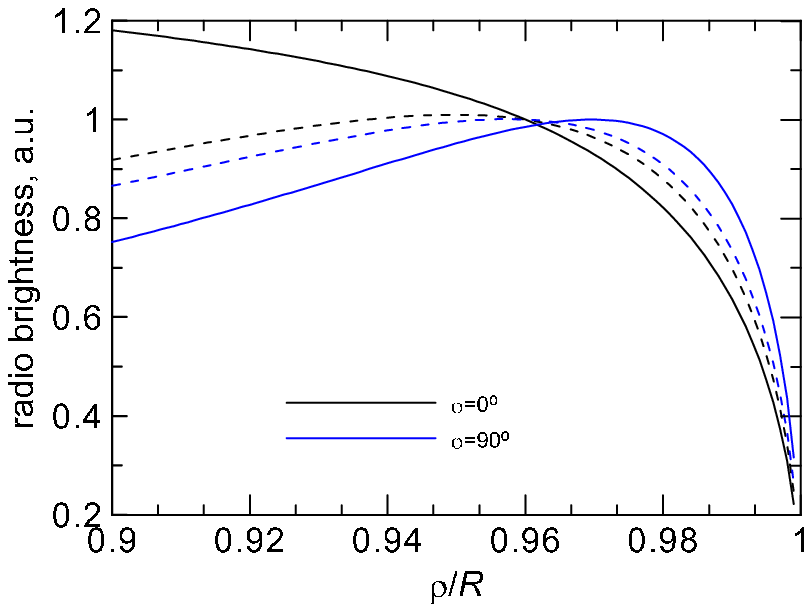} 
\caption{Radial profile of the radio brightness $S\rs{\varrho}$ (solid lines) and its approximation (\ref{xmaps:app-d:i}) (dashed lines) for azimuth $\varphi=0^\mathrm{o}$ (black lines) and $\varphi=90^\mathrm{o}$ (blue lines), $\phi\rs{o}=90^\mathrm{o}$. For smaller aspects, $\phi\rs{o}<90^\mathrm{o}$, the approximation agree better with the numerical profiles.
              } 
\label{xmaps:app-d-R}
\end{figure}

\section[]{Approximate formula for 
the radio surface brightness in Sedov SNR}
\label{xmaps:app5}

An analytic approximation for the azimuthal variation of the radio surface brightness of Sedov SNR (Paper I) may be extended to allow also for a description of the radial variation close to the forward shock. Namely, the correction consists in a factor $I\rs{r}$:
\begin{equation}
 S\rs{r}(\varphi,\bar\varrho)\propto
 \varsigma(\varphi)\sigma\rs{B}(\varphi)^{(s+1)/2}
 I\rs{r}(\bar\varrho)
 \label{xmaps:app-d:s}
\end{equation}
where $I\rs{r}$ 
is the same as for the X-ray approximation (\ref{xmaps:Ix}).
Accuracy of this approximation for the radial profile of brightness is demonstrated on Fig.~\ref{xmaps:app-d-R} and on Fig.~2 in Paper I for the azimuthal profiles. $I\rs{r}$ varies with azimuth less than $10\%$ (cf. e.g. black and blue dashed lines on Fig.~\ref{xmaps:app-d-R}). This variation is due only to $\beta(\Theta\rs{o})$. Thus, $\beta$ may be taken constant with a good choice $\beta/2=1$ (see also Appendix~\ref{xmaps:app1}). 

The smaller $\phi\rs{o}$, the smaller differences between the radial profiles for azimuth $\varphi=0^\mathrm{o}$ and $90^\mathrm{o}$ (black and blue solid lines approach one another with decrease of the aspect angle).

\label{lastpage}

\begin{thebibliography}{99}
 \bibitem[{Acero} {at al.}(2010)]{HESS-sn1006-2010}
  Acero F., et al., 2010, A\&A, 516, A62 
 \bibitem[{Ballet}(2006)]{Ballet-2006}
  {Ballet J. 2006, Adv. Space Res., 37, 1902}
 \bibitem[{Berezhko} {et al.}(2003)]{Ber-Volk-2003-mf}
  Berezhko, E. G., Ksenofontov, L. T. \& V\"olk, H. J. 2003, A\&A, 412, L11 
 \bibitem[{{Berezhko} \& {V\"olk}(2004)}]{Ber-Volk-2004-mf}
  Berezhko E. G. \& V\"olk H. J., 2004, A\&A, 419, L27 
 \bibitem[{{Cassam-Chena\"i} {et al.}(2005)}]{Decours-2005-prof} 
  Cassam-Chena\"i G., Decourchelle A., Ballet J., Ellison D. C., 2005, A\&A, 443, 955
 \bibitem[{{Cassam-Chena{\"i}} {et~al.}(2008){Cassam-Chena{\"i}}, {Hughes},
          {Reynoso}, {Badenes}, \& {Moffett}}]{chr08}
  {Cassam-Chena{\"i}} G., {Hughes} J.~P., {Reynoso} E.~M., {Badenes} C., \&
  {Moffett} D., 2008, ApJ, 680, 1180
 \bibitem[{Ellison} {et al.}(2000)]{Ell-et-al-2000} 
  {Ellison, D. C., Berezhko, E. G., \& Baring, M. G. 2000, ApJ, 540, 292}
 \bibitem[{Ellison} {et al.}(2001)]{Ell-et-al-2001} 
  {Ellison, D. C., Slane, P., \& Gaensler, B. M. 2001, ApJ, 563, 191}
 \bibitem[{{Ellison} \& {Cassam-Chena\"i}(2005)}]{Ell-Cassam2005-profiles} 
  Ellison D. \& Cassam-Chena\"i G., 2005, ApJ, 632, 920
 \bibitem[{{Hnatyk} \& {Petruk}(1999)}]{1999A&A...344..295H}
  {Hnatyk} B. \& {Petruk} O., 1999, A\&A, 344, 295
 \bibitem[{Fryxell} {et al.}(2000)]{for00}
  Fryxell B., Olson K., Ricker P. et al., 2000, ApJS, 131, 273
 \bibitem[{{Fulbright} \& {Reynolds}(1990)}]{reyn-fulbr-90} 
  {Fulbright} M.~S. \& {Reynolds} S.~P., 1990, \ApJ, 357, 591  
 \bibitem[{Kang} \& {Ryu}(2010)]{Kang-Ryu-2010}
  {Kang, H. \& Ryu, D. 2010, ApJ, 721, 886}
 \bibitem[{Lazendic} {et al.}(2004)]{Lazendic-et-al-2004} 
  {Lazendic, J. S., Slane, P. O., Gaensler, B. M., et al. 2004, ApJ, 602, 271}
 \bibitem[{{Long} {et al.}(2003)}]{Long-et-al-2003} 
  Long K. S., Reynolds S. P., Raymond J. C., et al. 2003, ApJ, 586, 1162
 \bibitem[{{Miceli {et al.}}(2009)}]{SN1006Marco} 
  Miceli M., Bocchino F., Iakubovskyi D., Orlando S., Telezhinsky I., 
  Kirsch M. G. F., Petruk O., Dubner G., Castelletti G., 2009, A\&A, 501, 239
 \bibitem[{{Lee} {et al.}(2008)}]{Ell2008-images} 
  Lee S.-H., Kamae T., Ellison D. C., 2008, \ApJ, 686, 325
 \bibitem[{{Morlino} {et al.}(2010)}]{Morlino-etal-2010}
  Morlino G.,Amato E., Blasi P., Caprioli D., 2010, MNRAS, 405, L21
 \bibitem[{{Orlando} {et al.}(2007)}]{Orletal07}
  {Orlando} S., {Bocchino} F., {Reale} F., {Peres} G., \& {Petruk} O., 2007, 
  A\&A, 470, 927
 \bibitem[{{Orlando} {et al.}(2010)}]{Orletal10}
	{Orlando S. Petruk O., Bocchino F. \& Miceli M. 2010, A\&A, accepted [astro-ph.HE/1011.1847]}
 \bibitem[{Parizot} {et al.}(2006)]{Parizot-et-al-2006}
  {Parizot E., Marcowith A., Ballet J., Gallant Y. A. 2006, A\&A, 453, 387}
 \bibitem[{{Petruk}(2006)}]{Pet06} 
  {Petruk O. 2006, A\&A, 460, 375}
 \bibitem[{{Petruk}(2009)}]{Pet08IC} 
  Petruk O., 2009, A\&A, 499, 643
 \bibitem[{{Petruk} {et al.}(2009a)}]{thetak} 
  Petruk, O., Beshley, V., Bocchino, F., \& Orlando, S., 2009b, MNRAS, 395, 1467 (Paper II)
 \bibitem[{{Petruk} {et al.}(2009b)}]{Petetal09icp} 
  Petruk O., Bocchino F., Miceli M., Dubner G., Castelletti G., Orlando S., Iakubovskyi D., Telezhinsky I., 2009b, MNRAS, 399, 157 
 \bibitem[{{Petruk} {et al.}(2009c)}]{pet-SN1006mf} 
  Petruk O., Dubner G., Castelletti G., Iakubovskyi D., Kirsch M., Miceli M., 
     Orlando S., Telezhinsky I., 2009, MNRAS, 393, 1034 (Paper I)
 \bibitem[{{Reynolds}(1996)}]{Reyn96}
  Reynolds S.P. 1996, ApJ, 459, L13
 \bibitem[{{Reynolds}(1998)}]{Reyn-98} 
  Reynolds S. P., 1998, ApJ, 493, 375
 \bibitem[{{Reynolds}(2004)}]{Reyn-04} 
  Reynolds S. P., 2004, Adv. Sp. Res., 33, 461
 \bibitem[{Sedov}(1959)]{Sedov-59}
 Sedov L.I., 1959, {\it Similarity and Dimensional Methods in Mechanics} (New York,
 Academic Press) 
 \bibitem[{Schure} {et al.}(2010)]{Schure-et-al-2010}
  {Schure K. M., Achterberg A., Keppens R., Vink J. 2010, MNRAS, 406, 2633}
 \bibitem[{Uchiyama} {et al.}(2003)]{Uchiyama-et-al-2003} 
  {Uchiyama, Y., Aharonian, F. A., \& Takahashi, T. 2003, A\&A, 400, 567}
 \bibitem[{Vink} {et al.}(2006)]{Vink-et-al-2006}
  {Vink J., Bleeker J., van der Heyden K., Bykov A., Bamba A., Yamazaki R. 2006, ApJ, 648, L33}
 \bibitem[{Zirakashvili} \& {Aharonian}(2007)]{Zirakashv-Aha-2007}
  {Zirakashvili V., Aharonian F., 2007, A\&A, 465, 695}
 \bibitem[{Zirakashvili} \& {Aharonian}(2010)]{Zirakashv-Aha-2010}
  Zirakashvili V., Aharonian F., 2010, ApJ, 708, 965
\end{thebibliography}
\end{document}